\documentclass[namedreferences]{solarphysics}
\usepackage[optionalrh,natbib]{spr-sola-addons} 

\usepackage[pdfborder={0 0 0 },urlcolor=blue,breaklinks]{hyperref}
\usepackage{graphicx}
\usepackage[usenames]{color}

\def\spose#1{\hbox to 0pt{#1\hss}}
\def\simlt{\mathrel{\spose{\lower 3pt\hbox{$\mathchar"218$}}
     \raise 2.0pt\hbox{$\mathchar"13C$}}}
\def\simgt{\mathrel{\spose{\lower 3pt\hbox{$\mathchar"218$}}
     \raise 2.0pt\hbox{$\mathchar"13E$}}}
\def\lsim{\rlap{$<$}{\lower 1.0ex\hbox{$\sim$}}}
\def\gsim{\rlap{$>$}{\lower 1.0ex\hbox{$\sim$}}}

\long\def\symbolfootnote[#1]#2{\begingroup%
\def\thefootnote{\fnsymbol{footnote}}\footnote[#1]{#2}\endgroup}

\begin{document}

\begin{article}

\begin{opening}

\title{What have we learned from helioseismology, what have we really
  learned, and what do we aspire to learn?\symbolfootnote[1]{Invited Article}}

\author{Douglas \surname{Gough} }
\institute{Institute of Astronomy, Madingley Road,
  Cambridge, CB3 0HA; Department of Applied Mathematics and
  Theoretical Physics, Centre for Mathematical Sciences, Wilberforce
  Road, Cambridge, CB3 0WA, UK \email{douglas@ast.cam.ac.uk};\\ 
  Physics Department, Stanford University, CA 94305, USA}

\runningauthor{D.O. Gough}
\runningtitle{Learning from Helioseismology}

\date{}

\begin{abstract}
Helioseismology has been widely acclaimed as having been a great success: it appears to have answered nearly all the questions that we originally asked,  some with unexpectedly high precision.  We have learned how the sound speed and matter density vary throughout almost all of the solar interior --  which not so very long ago was generally considered to be impossible -- we have learned how the Sun rotates, and we have a beautiful  picture, on a coffee cup, of the thermal stratification of a sunspot, and also an indication of the material flow around it.  We have tried, with some  success at times, to apply our findings to issues of broader relevance: the test of the General Theory of Relativity via planetary orbit precession  (now almost forgotten because the issue has convincingly been closed, albeit no doubt temporarily),  the solar neutrino problem, the manner of the transport of energy from the centre to the  surface of the Sun, the mechanisms of angular-momentum redistribution, and the workings of the solar dynamo.  The first two were of general interest to the broad scientific community beyond astronomy, and were, quite rightly, principally responsible for our acclaimed success; the others are still in a state of flux.
\end{abstract}


\end{opening}

\section{Prelude}
In the early heady days of helioseismology, the new techniques of inference, coupled with pertinent observations that had been stimulated by them, went hand-in-hand with consequent scientific discovery, and helioseismology was not unnaturally perceived to be almost a branch of science.  Now, in its relative maturity, it is, or at least it should be, relegated to what it really is: a very valuable technique for drawing scientific inference.  The basic  physics of seismic oscillations is not new, is fundamentally quite simple, and is well understood; and the principles of inference from the observations should be straightforward to comprehend, even though the technicalities of putting them into practice may for some seem to be rather complicated.  Therefore the robust raw conclusions are genuinely secure, more so than the broader issues to which they are intended to be applied.  However, it is incumbent upon us to make the distinction between the inferences that really have been drawn reliably and the further wider inferences that might subsequently be, or have been, drawn, often with the aid of supplementary, possibly less secure, maybe non-seismic, information, and even, maybe, (sometimes unstated) surmise.  Only if such distinction is made clear can the contributions of our subject to science be reaped to the full.  Unfortunately, that aspiration has not always been achieved in the past, and misinformation has sometimes sullied the waters.

It takes only a brief scrutiny of the equations describing the structure and dynamical evolution of the Sun (it is not quite so brief to derive them) and the equations governing the low-amplitude seismic modes of oscillation to appreciate what broadly can, at least in principle, be reliably inferred.  Anything further must depend on other criteria, such as general physical argument beyond seismology, traditional astronomical observation, or even prejudice.  It is obligatory to be explicit about how such additional constraints are applied.  The subject has advanced to a new level of sophistication; we are now trying to probe seismically (and otherwise) almost inaccessible aspects of the physics of the Sun, and the techniques for unravelling them are becoming more and more intricate, beyond the point at which most scientists wish to tread.  There must necessarily be an increased trust in our findings, and it is our responsibility not to betray it.  Many of the broader scientific community want to use our results in their research; for that they need to know not only the limitations of our inferences, and the caveats upon which they are based, but also which aspects of what we seismologists tell them can really be trusted.

Much of the emphasis of \textit{Solar Dynamics Observatory} seismology concerns the workings of the convection zone.  We want to know what controls the solar cycle, how magnetic field is amplified, modulated, and then suppressed, how sunspots are formed and destroyed -- and what determines their lifespan. We want to know  the geometry of at least the larger scales of convective motion, and how, beneath the seen superficial layers of the Sun, the processes that control the total radiative output are modulated.  At least some of us want to understand how all these matters influence our procedures for inferring the gross properties of the Sun, and how they impinge on our broader ideas of the evolution of the Sun in particular, and of stars in general.  Addressing such delicate issues with confidence may now seem an almost impossible task to us older scientists who have lived through the years of stumbling through the darkness,  having finally emerged to bathe in the secure light illuminating the minute arena of knowledge that we have been instrumental in uncovering.  It is now up to the younger community to proceed likewise: to grasp at the edge of our perception with initially insecure ideas, fully appreciating the uncertainty, of course; then moulding and strengthening them into a new body of secure scientific knowledge.

\section{Introduction}     
Once the potential of solar oscillations to map the interior of the
Sun was recognized \citep{JCDDOGheliology1976} there were two obvious serious
issues that immediately appeared accessible to resolution: the spherically symmetric component of the
hydrostatic stratification and the internal angular velocity.  The first of
these was needed for investigating what has been called the solar
neutrino problem; the second concerned the centrifugally induced
oblateness of the Sun's gravitational equipotentials, and how that impinged
upon an important test of theories of gravity, General Relativity in
particular, via the precession of planetary orbits.  These were the
two most widely discussed issues in heliophysics at the time.

The stratification was originally addressed with the help of
theoretical models.  \citet{deubner1975A&A} had published the first well
resolved $k$--$\omega$ spectrum, and \citet{AndoOsaki1975PASJ} had shown that
theoretical eigenfrequencies of p modes trapped in the outer layers of
a solar model envelope  were in quite good, but not perfect,
agreement with Deubner's observations.
What was required for bringing the theory more closely into line with observation
was first estimated from the properties of the eigenfrequencies of a simple
polytropic representation of the outer layers of the convection
zone as a means of calibrating solar models \citep{DOG1977Nice}; 
the conclusion was that the convection zone must be about 200\,Mm deep,
some 50\,Mm deeper than the favoured value of the time.  Basically, that conclusion was drawn 
from relating theoretical eigenfrequencies to the jump in the adiabatic ``constant''
$p/\rho^{\gamma_1}$, which is closely related to specific entropy, across the thin superadiabatic convective boundary layer.  It was subsequently supported by more realistic, numerical, computations by \citet{rkuejrdepthconvzone1977ApJ}.  The principal implication of that result 
was that, according to complete solar models, a deeper convection zone implied
greater helium and heavy-element abundances, a hotter, more centrally
condensed, core, and a higher neutrino flux, thereby exacerbating the
solar neutrino problem  \citep[\textit{e.g.}][]{abrahamiben1971ApJ, bahcallulrich1971ApJ}.
Indeed, one is tempted to speculate that the modellers in the past had adjusted
the defining parameters of their models to minimize the theoretical
neutrino flux, and that that had prejudiced Ando and Osaki's calculations,
although without repeating the calculations oneself (and maybe even if
one did) one cannot be sure.

There was a great deal of healthy mistrust in the models at that
time.  The concern was that a seismological measurement -- I should really say
estimate -- of merely the upper boundary layer of the convection zone,
extending only a minute fraction of the solar radius beneath the
photosphere, could hardly be a robust indicator of conditions inside
the energy-generating core.  Therefore, as soon as low-degree data
became available  \citep{isaaketaltucson1980LNP, grecfossatpomerantz1980Natur, fossatgrecpomerantz1981SoPh} it 
became possible to use seismic indicators of more global properties;
first the so-called large frequency separation \citep{dirtymodels1979A&A....73..121C,dirtymodelscorrigendum1979A&A....73..121C,JCDDOGTucson1980LNP}, which measures the sound travel time from the centre of the Sun to the seismic surface\footnote{I consider  the seismic surface $r=R$ of the Sun (assumed here to be spherically symmetrical) to be the radius at which $c^2$, regarded as a function of $r$,  or $c$, regarded as a function of acoustic radius  
$\tau(r)=\int{c^{-1}{\rm d}r}$  -- both of which are close to being linear functions in the outer adiabatically stratified layers of the convection zone \citep{effluentpulsation1990ApJ,DOGIlidio2001MNRAS.322.473L} -- extrapolate to zero.  In the Sun, according to Model S of \citet{jcdetal1996Sci}, it lies about 1000\,km above the photosphere, the precise value depending on exactly how the extrapolation is carried out.  
There is nothing special about the structure of the actual atmosphere in its vicinity, which is well inside the outer evanescent zone of most of the seismic modes and therefore has little significant  influence on the dynamics.  Instead, it 
acts simply as a (virtual) singularity in the acoustic wave equation, providing a convenient parametrization of conditions (well below the photosphere) in the vicinity of the upper turning points of the modes.  Put another way, it provides a convenient fiducial location with respect to which the acoustic phase in the propagating zone beneath is related. Unlike the  photosphere, which has no acoustic significance, it shares a relation with the deeper solar interior that is robust, and is insensitive to the non-seismic, thermal and radiative, properties of the outer convective boundary layer, whose structure changes with the solar cycle \citep{antiabasuradiuschanges2004ESASP.559..301A,solarcyclefreqchange2004ApJ.600.464D,solarcyclefreqchange2005ApJ.625.548D,solarcycleradiuschanges2005ApJ.633L.149L,solarcycleradiuschanges2007ApJ.658L.135L}.  In contrast to other, non-seismic,  radii  \citep[\textit{cf.}][]{bahcallulrich1988RvMP.60.297B}, it provides a stable outer limit to the effective total acoustic-radius integral $\tau(R)$, which determines the large frequency separation; in Model S it is some 200 seconds or so greater than the actual acoustic radius of the photosphere.}
\citep[\textit{e.g.}][]{vandakurov1967AZh.44.786V,tassoulasymptotics1980ApJS.43.469T,DOGEBK1986}, and later the small frequency separation \citep{DOGprotosolarY1983}, which is a direct 
indicator of conditions in the core \citep[\textit{e.g.}][]{DOG1983PhysBull, DOGEBK1986}.  
Unfortunately it was not possible to make all of 
the theoretical frequencies agree with the data, as was clear from an
indiscriminant attempt to fit only the low-degree frequencies with
whole-disc observations \citep{JCDDOG1981A&A}: two solar models seemed to
be  favoured, one with a low initial helium abundance: $Y_0 = 0.18$ (coupled with a 
correspondingly low heavy-element abundance $Z=0.003$ and a low neutrino flux, although not low enough to
reproduce the neutrino detection rate), the other large: $Y_0=0.27$ and $Z_0=0.026$.
The high-$Y$ model fitted rather better, and one was tempted to
prefer it, especially because its high-degree mode frequencies were closer to 
Deubner's observations.  Moreover, the low-$Y$ model had a helium
abundance below what was thought to have been produced in the Big
Bang (\textit{cf.} footnote 8), which would call for some contrived explaining.

The two models required different identifications of the orders [$n$] of
the modes; unfortunately the orders were so high, and the fit so poor,
that reliable extrapolation to $n=1$ was not possible.  It was not until \citet{duvallharveybridgeingap1983Natur} observed the frequencies of modes of
intermediate degree that a secure connection between modes of low and high
degree could be made \citep{DOGBridgeingap1983Natur}.  The orders of the latter are  determinable because the frequencies of 
high-degree f modes, with which is associated $n=0$, are essentially
independent of the structure of the Sun \citep{DOGledouxboulderreview1982pccv.conf..117G}.  
The high-$Y$ alternative was thereby confirmed.  

Thus we had learned that the solar neutrino problem was almost
certainly not resolvable by adjusting solar models, and must be a
matter for nuclear or particle physics.

\begin{figure*}
\centering
     \begin{center}
       \includegraphics[width=8.0cm]{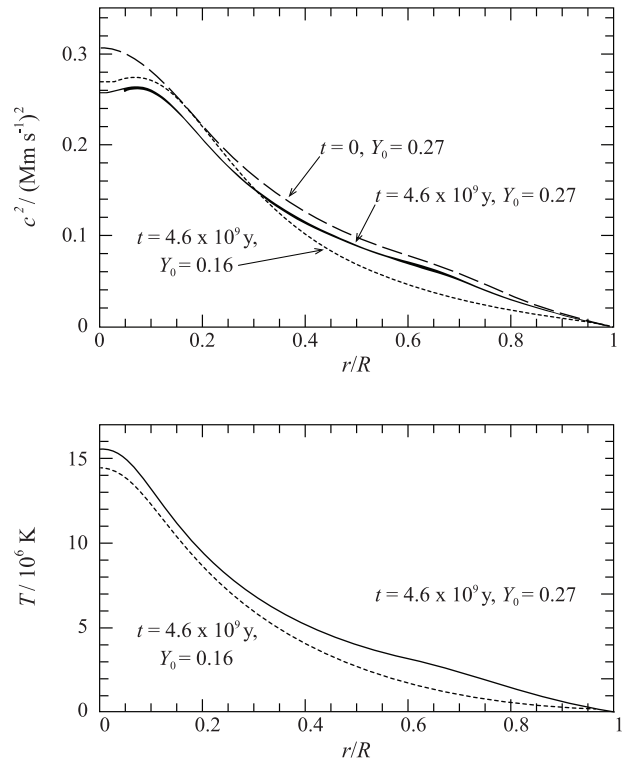}
     \end{center}
\caption{Upper panel: The continuous curve extending over the entire range of $r/R$ is the square of the sound speed [$c^2$]  in the standard solar 
model S  of \citet{jcdetal1996Sci}, which was computed with initial hydrogen and heavy-element abundances $X_0 = 0.7091,\;  Z_0 = 0.0196$.  The square of the sound speed in the Sun is also plotted (where $r/R>0.1$) as a continuous curve; the difference between the two can be barely discerned by the variation in the thickness of the apparently single curve (but see  Figures \ref{fig3} and \ref{fig4}).  The dip in $c^2$ at the centre of the Sun is a result of the augmentation of $\mu$ by nuclear transmutation, which increases with time on the main sequence (and which can therefore  be used as a diagnostic of main-sequence age).  The dashed curve is $c^2$ on the zero-age main sequence, and, except near the surface where abundant elements undergo ionization, is (locally) proportional to temperature.    The dotted curve is  $c^2$ in a  model with an initial heavy-element
abundance  $Z=0.001$, but continuously contaminated at the surface at  such a constant rate 
 as to have a current heavy-element abundance $Z_{\rm{s}}=0.02$ 
in the convection zone today \citep{dirtymodels1979A&A....73..121C}.  
Lower panel:  Temperature [$T$] through the two present-day solar models.}\label{fig1}
\end{figure*}

Had we really learned that?  Perhaps not yet.  At the time there were
still very serious doubts about the solar models, for they depended on
many unproven assumptions, some of which are listed in Table 1.  There
had already been, and there were yet to be, many models computed in
which some of these assumptions were relaxed in the hope of yielding lower
neutrino fluxes (although none provided a satisfactory reconciliation
of theory with observation).  At the very least, a secure
representation of the stratification throughout the Sun was surely required.
That was soon to be provided from inversion analyses of frequency data
from Duvall and Harvey \citep{speedofsoundJCDetal1985Natur, JCDDOGMJTdifferentialc2inversions1989MNRAS, DziemPamykSienksolarinversion1990MNRAS}, from which it was possible to infer the
(spherically averaged) sound speed [$c(r)$] throughout almost all of the
Sun.  Details of the core were not yet within reach (and even today
there is considerable uncertainty), but elsewhere the sound speed was
essentially the same as that of Christensen-Dalsgaard and Gough's (1981)
high-$Y$ model.  It was also possible to see the base of the
convection zone.  I recall one Friday morning (the second Friday of
January, 1984,  a day on which I was due
to deliver a lecture on my findings to the Royal Astronomical
Society);  it was about 4:30 in the morning -- one could compute seriously only at night in
those days -- when I obtained my first plot of $c^2(r)$ that extended
beneath the convection zone; I plotted $c^2$ and not $c$ because it is 
related closely to temperature $T$ -- the equation of state in the solar interior is reasonably well 
approximated by the perfect-gas law, for which $c^2 \propto T/\mu$,  where $\mu$ 
is the ``mean molecular mass''.  A more modern
(and rather better) version of what I obtained is illustrated in
Figure \ref{fig1}\footnote{Included for comparison is one of the models with low $Y$ that was 
used in the original calibration with low-degree  modes by \citet{JCDDOG1981A&A}.  
The qualitative differences between the models can be appreciated by realizing first that the radiative envelopes are roughly polytropic (with index ~3.5), and that it is adequate to approximate the equation of state by the perfect-gas law. Then it is clear that the magnitude of $T(r)$ (and $\rho$) must be greater in the higher-$Y$ model, because the total energy generation rate, which is an increasing function of $X$, $\rho$, and $T$, is the same for the two models.  Polytropic scaling 
\citep[\textit{e.g.}][]{1990stromgren} indicates that $Z$ is a steeply decreasing function of $X$, so $Z$ is much greater in the higher-$Y$ model.  In addition, the polytropic radius scale is greater for the higher-$Y$ model, as is evident from the Figure  \ref{fig1} by imagining an extrapolation of the functional form of $T(r)$ outwards from the radiative zone.  Consequently the convection zone, which steepens the gradient, has more truncating from the radiative structure to perform in order to maintain the observed photospheric radius, and is therefore deeper.  An additional scaling in magnitude is 
required to convert $T$ to $c^2 \propto T/\mu$, raising  the dotted curve (corresponding to the 
model with the lower $Y$) relative to the continuous curve.   It also depresses both curves near the centre of the star, in the energy-generating core where $\mu$ has been augmented by nuclear transmutation, providing a diagnostic of main-sequence age.}.
The location of the base
of the convection zone is evident as a near discontinuity of the
second derivative, as one can see more easily by holding the page
almost in one's line of sight and looking along the curve.  I
confirmed the location by having a research student, the only person left
in the computer room at that hour, repeat the exercise.  I went
home, slept a couple of hours, and then took a train to London.  When I
showed the plot at the RAS \citep{DOGRASc2inversiontalk1984} and pointed out the 
discontinuity, the audience was incredulous, even though I stood almost in the plane of
the screen to reassure them.  I then recounted the confirmation by the
research student, and there was a sudden release of tension in the
audience when I told them that the student was a cosmologist, for then they
appreciated that such a student must undoubtedly have provided an unbiased
opinion on such a matter. 

The result was compared with theoretical solar models \citep{speedofsoundJCDetal1985Natur}, 
and revealed a characteristic discrepancy (much larger than that evident
in the more modern comparisons, such as those illustrated in Figures \ref{fig3}
and \ref{fig4}) immediately beneath the base of the convection zone.
It was suggested that the discrepancy could have been caused by an
error in the opacity (values of opacity in those days were available in tables provided by the Los Alamos National Laboratory in the USA). Such
suggestions, in a rather broad sense, had been made in the past, 
for example by \citet{simon1982ApJ...260L..87S} in response to a failure  to reproduce simultaneously observations relating to stellar evolution and observations of pulsational characteristics of classical variable stars.   But here one
was able to state quite precisely the thermodynamical conditions under which the
opacity was in error, and the approximate magnitude (and sign) of the error.
After some persuasion, Carlos Iglesias and Forrest Rogers at Livermore computed by moonlight the
opacity for a few judiciously chosen values of the state variables $\rho$ and $T$ (and a plausibly appropriate chemical composition) using an independently written computer programme; 
they not only confirmed the helioseismic inference but also showed that
the discrepancy was even greater at lower temperatures -- of no concern
for the Sun because it is well inside the adiabatically stratified region
of the convection zone where radiative energy transfer is negligible, but of great significance in reconciling with 
observation the theory of some classes of intrinsically variable stars
such as $\beta$~Cephei and slowly pulsating B stars \citep{coxetalOPALbetacepheipulsation1992ApJ.393.272C,kiriakidisetalOPALbetacepheipulsations1992MNRAS.255P.1K,moskalikdziembowskibetacephei1992A&A,DziembowskiSPBinstabI1993MNRAS.262.204D,DziemboskiSPBinstabII1993MNRAS.265..588D,DziembowskietalSPBstars1994IAUS} and double-mode Cepheids \citep{moskaliketalcepheidperiodratios1992ApJ.385.685M}.
Largely as a result of efforts by Werner D\"{a}ppen, scientists who were involved with the  Los Alamos opacity computations were 
brought into the same room as the Livermore scientists, and errors
in the Los Alamos calculations were identified.  This led to the Lawrence Livermore National Laboratory  providing the funds for more extensive opacity computations for astrophysical use 
\citep{iglesiasrogerscepheidopacity1990ApJ, Iglesiasrogersopacity1991ApJ, updatedopacity1996ApJ.464.943I,rogersnayfonov_EoS_2002ApJ...576.1064R}.  Thus we see the first
example of helioseismology contributing directly to microscopic physics (if one considers the neutrino issue to be indirect).  There
have been further contributions, but I postpone discussion of those until
later.  

I emphasize that an important consequence of the sound-speed inversion
was that it convincingly ruled out the low-$Y$ models, reaffirming that
the cause of the neutrino deficit was not in solar modelling.  That conclusion was not fully appreciated by many of the solar modellers of the time, for they had not yet understood 
the power of helioseismological analysis.

\begin{figure*}
\centering
     \begin{center}
       \includegraphics[width=8.0cm]{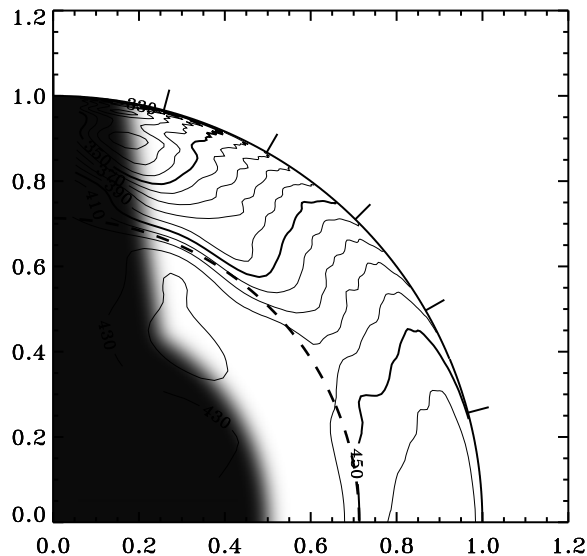}
     \end{center}
\caption{Optimally localized  averages of the North\,--\,South symmetric component of the angular velocity [$\Omega$] of the Sun, depicted as a contour plot in a quadrant.  Some contours are labelled [nHz], and, for clarity, every fifth contour is drawn bold; the contour separation is 10\,nHz. The 
outer quarter circle denotes the surface of the Sun,  the dashed quarter circle indicates the base of convection zone, and the tick marks at the Sun's surface are drawn at latitudes 15, 30, 45, 60, and $75^{\circ}$.  The Equator is the horizontal (relative to the page) axis and the Pole the vertical, each labelled with values of  $x=r/R$. The shaded area indicates the region in the Sun where no reliable inference could be made from the data available   \citep[from][]{schouetalrotation1998ApJ}.  There has been little significant improvement of this inference, save for a demonstration that the core is not rotating rapidly \citep{elsworthetalrotation1995Natur.376..669E,chaplinetalrotation1999MNRAS.308..405C}; however, there have been studies of temporal variation associated with the solar cycle 
\citep[\textit{e.g.}][]{vorontsovetaltorsionalosc2002Sci,
basuantiarotationvariation2003ApJ...585..553B,
howeetalrotationvariationI2006SoPh..235....1H,
howeetalrotationvariationII2006ApJ...649.1155H,
antiachitregoughsolarKE2008A&A}.}\label{fig2}
\end{figure*}

I turn now to the angular velocity [$\Omega$], which is measured by
the odd (with respect to azimuthal order $m$) component of degeneracy
splitting caused by advection and Coriolis acceleration.  Strictly speaking,
one needs the sound-speed inversion first, to establish the
hydrostatic stratification with respect to which the splitting kernels are
computed.  However, the angular velocity was actually first obtained by using a
theoretical solar model before the full sound-speed inversion had been carried out \citep{Naturerotation1984Natur}, partly,
perhaps, because it was easier to perform with the data in hand, yet also with the knowledge that  the error from which that procedure suffered was much less than that resulting from the errors in the  measurements of the degeneracy splitting.   The
result was a big surprise: 

There had been many prior discussions of
how much faster than the photosphere the solar core must be rotating (due to spin-down), the
most widely publicized being the discussions resulting from the
measurements of the apparent oblateness of the solar surface by \citet{DickeGold1967PhRvL,  dickegoldenberg1974ApJS}.  The oblateness measurements had been made in
the hope of confirming Dicke's theory of gravitation  \citep{BransDicke1961}\footnote{The gravitational attraction associated with the energy density $-GM/r$ of the gravitational field surrounding the Sun, absent in Newton's theory, causes the total gravitational attraction to increase: very roughly speaking, as a result of energy conservation the apparent gravitational mass of a planet at distance $r$ from the Sun, in a flat representation of space,  is augmented by approximately $GM/c^2r$ per unit mass above what it would have appeared to have been at infinity;  $M$ is the mass of the Sun, and here $c$ is the speed of light. Similarly, the energy, hence the frequency, of a photon is multiplied by a factor $\Gamma = 1+GM/c^2r$ -- that causes the familiar gravitational redshift.  Consequently, the orbit equation is modified simply by multiplying the Newtonian gravitational force on the planet by $\Gamma^3$.  After linearization and rewriting $M$ in terms of the orbital specific angular momentum $h=\sqrt(GMr)$, valid for nearly circular orbits, the effective attractive force becomes $-(1+3h^2/c^2r^2)GM/r^2$; it increases with increasing  proximity more rapidly than Newton's  inverse square.
It is easy to see also that the gravitational field in the equatorial plane of a rotating (oblate) axisymmetric self-gravitating body (like the Sun) also increases with decreasing distance faster than the field around a corresponding spherically symmetrical body:  the act of flattening a spherical body takes equal amounts of material from the poles towards the near and the far sides of the equator,  the increase in gravitational attraction by closer nearside matter exceeding the lesser decrease by farside matter. Therefore the net gravitational attraction is increased, by an amount which inceases as $r$ decreases and the shape of the Sun becomes more apparent.  The force is given approximately by $-(1+\frac{3}{2}J_2R^2/r^2)GM/r^2$; $J_2$ is the quadrupole moment.  In both cases, therefore, a planet is drawn towards the Sun increasingly strongly with increasing proximity than it would have been in an inverse-square field.  Conserving its angular momentum, it is thereby caused to rotate through a greater angle near perihelion because its orbital angular velocity is augmented, distorting an otherwise approximately elliptical bound Newtonian orbit in such a manner as to appear to make it simply precess in the same direction as the angular velocity of the planet.  (Near aphelion 
the oppositely directed contribution to the precession is lesser, therefore too small to annul the contribution from near perihelion.)  
Use of planetary (or spacecraft) orbital precession rates  to calibrate the relativistically induced deviation from the inverse-square gravitational field surrounding the Sun therefore requires one to know the contribution from the distortion from spherical symmetry of the mass distribution in the Sun, such as is produced by centrifugal acceleration due to rotation.  The precession rate is most easily calculated by perturbation theory \citep[\textit{e.g.}][]{Ramseydynamics}.}: put naively, in the Brans--Dicke theory 
Newton's gravitational constant [$G$] was regarded as a field satisfying
a wave equation that couples to the matter, thereby relaxing the gravitational force field 
from the relative rigidity that is imposed when $G$ is  held constant, and hence
reducing the rate of precession of planetary orbits from the value
predicted by General Relativity; the fact that General Relativity
predicts essentially the correct precession with a spherical Sun was
regarded by Dicke as a fortuitous coincidence, and that in reality the
shortfall predicted by his theory (which is not an absolute prediction, but
depends on an unknown coupling constant) is made up by the
oblateness of the Sun's gravitational field produced by a rapidly
rotating interior.  The oblateness measurements would calibrate the 
coupling constant.  A debate ensued concerning two matters: the
relation between the oblateness in surface brightness and the
oblateness of the gravitational equipotentials  \citep[\textit{e.g.}][]{dicke1970ApJ}, and
the fluid dynamics of the solar interior concerning angular-momentum
transfer from centre to surface during spin-down  \citep[\textit{e.g.}][]{howardmoorespiegel67, dickespindown1967ApJ, brethertonspiegel1968ApJ}.   I shall enlarge on neither here,
despite the intrinsic interest of each, because even the early seismological analysis \citep{Naturerotation1984Natur} revealed that the Sun is unexpectedly rotating almost uniformly throughout its interior, except possibly in what might be its almost inaccessible core (Figure \ref{fig2}).  Subsequent observations of low-degree modes revealed that the
core rotates significantly no more rapidly than the envelope \citep{elsworthetalrotation1995Natur.376..669E,chaplinetalrotation1999MNRAS.308..405C}.  
From knowledge of $\Omega$ and the
density and pressure stratification, the multipole moments $J_{2k}$ of
the gravitational equipotentials can easily be computed \citep[\textit{e.g.}][]{DOGrotation1981MNRAS.196..731G, pijpers1998MNRAS.297L..76P}.  I maintain that
this is the most accurate way to determine $J_{2k}$, notwithstanding
some claims in the literature to the contrary.  The reason is partly
that other methods involve relating surface brightness asphericity
with gravity, which is not well understood, and partly because those
measures rely on only the small quadratic centrifugal force rather than the more 
robust, and unambiguous, seismic effects of advection and, to a much lesser degree,
Coriolis acceleration, both of which are predominantly linear in
$\Omega$ and are therefore much larger than the centrifugal term;  and, moreover,  they do not depend on relating brightness to gravity.  Moreover, the rotational distortion of the isobaric surfaces in the vicinity
of the photosphere is dominated by the direct effect of the local centrifugal
force, and exceeds the gravitational distortion by a factor of more than 20;
determining the latter from the apparent shape of the solar disc therefore
requires a subtraction of two measurements that differ by less than a mere 5
per cent. Even the most precise limb-shape measurements \citep[\textit{e.g.}][]{fivianetal2008,kuhnetalsolarshapeScience2012} are far from that goal \citep{dogscienceprospects2012}.

The outcome is consistent with General Relativity; indeed, the
miniscule contribution that $J_2$ does make to the precession of the orbit
of Mercury \citep[\textit{e.g.}][]{DOGrotation1982IrishAJ.15..118, Naturerotation1984Natur, brownetalrotation1989ApJ...343..526B, antiachitregoughsolarKE2008A&A} brought General
Relativity into somewhat closer agreement with observation than assuming the Sun to be spherically symmetrical.  The influence of higher moments is utterly negligible. The improvement
is marginal, but with future, more delicate, orbital analyses the helioseismological determination of  $J_2$  will evidently take on a crucial role.  The almost
uniform rotation of the radiative interior of the Sun begs the
question of how that can be, particularly because it interfaces with a
differentially rotating convection zone.  I shall return to that
matter later.

\section{Understanding Seismic Variables}
It is extremely important to understand what is actually inferred from
seismology, for only then can one draw reliable conclusions concerning
the Sun.  Seismic modes are essentially adiabatic.  They result from
forces -- pressure and gravity predominantly -- acting on matter with
inertia.  The dynamics therefore concerns only pressure [$p$] and matter 
density [$\rho$] which are related via gravity through hydrostatic
equilibrium, and the relation between perturbations to them under
adiabatic change, which is characterized by the first adiabatic
exponent [$\gamma_{1} = \left ( \partial \ln p / \partial \ln \rho
\right )_s$, the partial thermodynamic derivative being taken at
constant specific entropy [$s$] ].  The restoring force of acoustic modes
(p modes) is principally pressure, that of gravity modes (g modes,
including the fundamental f modes) is buoyancy.  So the modes can
provide information directly about only $p , \rho, $ and $\gamma_1$
(and, of course, any function of them).  These are the basic seismic
variables.  I should acknowledge that any magnetic field [$\bf{B}$] that
is present also contributes to the dynamics, and therefore is also a
seismic variable.  Unfortunately, its effect on the frequencies of
the modes appears to be indistinguishable  from that of an appropriate
variation in sound speed [$c$] -- also a seismic variable because $c^2=
\gamma_1 p/\rho$ -- which makes it difficult to unravel the two
\citep{egzdog1995ESASP}.  In principle one might be able to do so from the
eigenfunctions, which are configured differently by $c$ and $\bf{B}$,
although supplementation by non-seismic arguments is likely to be more productive. 

Leaving $\bf{B}$ aside for the time being, it should be appreciated
that the relation between $p$ and $\rho$ through hydrostatics does not depend directly on  
$\gamma_1$, so $p$ or $\rho$ (or any function of only them) and
$\gamma_1$ are structurally independent: it follows that although the relation between them,
which physically is given by the equation of state, can be determined by seismology 
along the thermodynamic $p$--$\rho$ path through the Sun, there is no remaining redundancy, 
so the veracity of the equation of state itself cannot therefore be probed by
seismology alone; in order to investigate the equation of state,
supplementary, non-seismic, information is required.

There are two approaches that one can take for drawing seismic
inference.  One is to adopt a parametrized model of the Sun, or
some aspect of it, and from it calculate whatever seismic properties
one wishes to compare with observation.  The comparison calibrates the
controlling parameters.   That was the procedure that I described in
my introduction for first estimating the depth of the convection zone,
and consequently the helium abundance.  The other approach is to
ignore explicit models (almost) entirely, and combine the data in such
a manner as to isolate certain properties of the seismic structure.
Typically that involves adopting, at least at first, a reference model
of the Sun, which it is hoped is sufficiently close to the Sun for
linearization in the small differences from it to be more-or-less valid;  that simplifies
the analysis enormously. Subsequent iteration can usually remove the
dependence on the reference model.  Nonlinear asymptotic methods have
also been used successfully to yield approximate inferences without
recourse to a reference model at all.  The most common procedures that
have been used to date are aimed at obtaining easily interpretable
representations of the basic seismic variables.  They are commonly called inversions. 

\begin{figure*}
\centering
     \begin{center}
       \includegraphics[width=11.0cm]{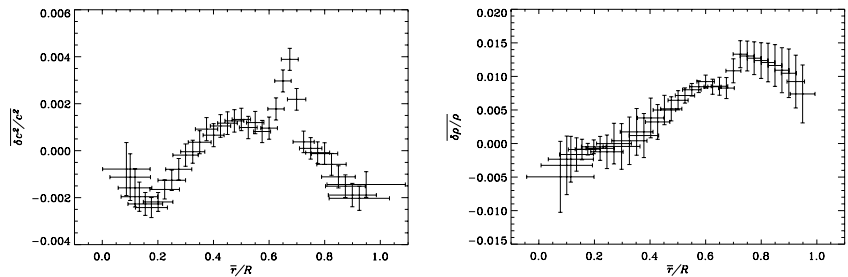}
     \end{center}
\caption{Optimally localized  averages of the relative differences of the squared sound speed [$c^2$] and  the density [$\rho$] in the Sun from those in Model S of Christensen-Dalsgaard \textit{et al.} (1996), computed by M. Takata  \citep{takatgough2001ESASP.464..543T} from MDI 360-day data and plotted against the centres [$\bar x = \bar r /R$] of the averaging kernels  [$A(x;\bar x)$]  (which here resemble Gaussian functions),   defined by $\bar x = \int x A^2 {\rm d}x\,/ \int  A^2 {\rm d}x$.   The length of each horizontal bar is {\it twice} the spread $s$  of the corresponding averaging kernel, defined as $s=12\int(x-\bar x)^2A^2{\rm d}x$ -- an averaging kernel $A$ that is well represented by a Gaussian function of variance $\Delta^2$ has spread approximately $1.7\Delta\approx0.72$\,FWHM; were it to be a top-hat function, its spread would be the full width, which is why $s$ has been so defined.  The vertical bars extend to $\pm 1$ standard deviation of the inversion errors, computed from the frequency errors quoted by the observers assuming them to be statistically independent; the errors in the averages are correlated \citep{errorcorrelation1996ApJ.459.779G,errorcorrelation1996MNRAS.281.1385H}.}\label{fig3}
\end{figure*}

\begin{figure*}
\centering
     \begin{center}
       \includegraphics[width=8.0cm]{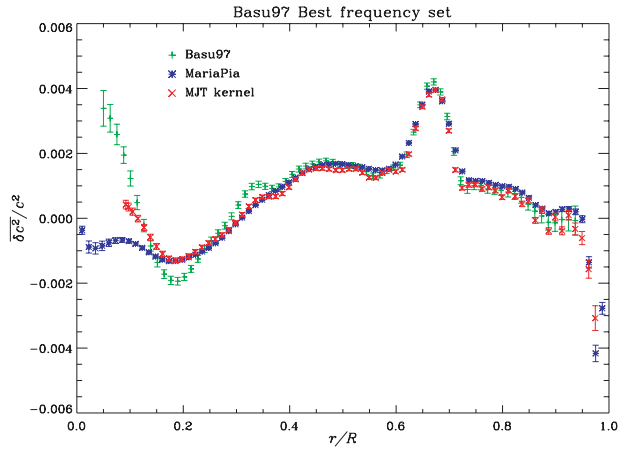}
     \end{center}
\caption{Several sets of optimally localized  averages of the relative difference between the squared sound speed in the Sun and that in Model S of Christensen-Dalsgaard \textit{et al.} (1996) computed from the same frequency data set, with kernels obtained by different inverters and by ignoring the 
contamination from other seismic variables.  The vertical bars extend to $\pm$ one standard deviation of the errors computed from the frequency errors quoted by the observers.  [The abrupt deviation close to the surface is a characteristic of having misrepresented the acoustic radius 
\citep[\textit{cf.}][]{MTDOGseismicradius2003ESASP.517..397T}.]}\label{fig4}
\end{figure*}

When linearization about a
reference model is carried out, the frequency differences between the Sun and the model 
can each be expressed as a sum of spatial averages of independent seismic variables.  A prudent procedure is then to seek suitable linear combinations of those averages which more easily inform 
one of whatever question one has chosen to pose.   If it is the value of a particular seismic variable that one wants to investigate, then the most
easily interpretable data combinations are those that represent averages of that variable with  weighting functions   
(called kernels in this context) that are highly localized with hardly
any sidelobes, and which at the same time are (almost) independent of any other variable, for then the result can be thought of as a blurred view of the 
variable of interest.  Consequently one  tries to tailor maximally
localized kernels, although the attempted maximization must be moderated by a requirement that the interference from extraneous seismic variables be kept low and that the
influence of data errors not  be excessive.   On the whole, increasing localization requires data combinations with coefficients of greater and greater magnitude, which increases the influence of 
random data errors.  The procedure to construct those kernels is now called
optimally localized averaging (OLA).  Just how low one demands the interference from other variables  to be, and how much the influence of data errors one considers to be acceptable, is a matter of personal choice, which probably explains why the errors in the averages displayed in Figures \ref{fig3} and \ref{fig4} differ. This kind of tailoring was the
first to be advocated for helioseismology \citep{DOG1978Cataniarotationinversion}, 
in contrast to the almost
universal opinion of geoseismologists of the time who objected to the
procedure on the ground that one might be tempted simply to draw a curve through
the averages and mistake that curve for the actual variable, rather
than the average that it actually is.  We helioseismolgists usually do
not make that mistake.  We explicitly draw a sequence of crosses, as
in Figure \ref{fig3}, the horizontal components representing the averaging
widths, the vertical bars representing the propagated standard errors
in the data.  Some geoseismologists are now changing their tune,
and are following suit.

An alternative procedure, common in geoseismology, is to seek a
putative ``solution'' that reproduces the data.  Typically, one first expresses the
seismic variables as linear combinations of a prescribed set of basis
functions, and then one chooses the coefficients in each combination
to match the data the most closely, again moderated (regularized) to
avoid excessive error magnification.  The reason geoseismologists used
to prefer that approach is that it might yield a curve that actually fits the data within the
estimated errors, which OLA might not (and is not explicitly
designed to).  If it is a linearized perturbation to a reference model that is being fitted to the data, the outcome can also be expressed as averages with localized components.  
However, the averaging kernels often have severe
sidelobes too, usually near the surface of the Sun, which renders
interpretation more difficult.  Moreover, the ``solution'', if it exists, is just one
of infinitely many that satisfy the constraints imposed by the data.
Usually, error-correlation information about the raw data is
unavailable, and the data are usually fit by (regularized) least squares
(RLS), regularization being accomplished typically by imposing some
criterion of smoothness on the ground that the data provide only
finite spatial resolution.  

If two seismic variables are involved, as
is the case for structure inversions (in the absence of rotation and a magnetic field), 
then two separate complementary
inversions must be carried out for OLA, each requiring the suppression
of the contribution to the data combination from the variable not
being sought.   With the help of some 
regularizing assumption, estimates of the seismic variables can then 
be made from their averages to enable 
one to gauge,  for each seismic average, the contribution to the corresponding data combination from the other seismic variable; in most of the published ``inversions'', that contribution is ignored.  In the case of  RLS frequency fitting, the two seismic variables 
can be represented simultaneously.

I do not here go into how the inversions are carried out.  That is not
necessary for understanding the results, provided that adequate
information about the averaging kernels is given.   Unfortunately, that
information is not always provided.  Precisely how the centres of the
optimally localized kernels were determined is rarely stated, although
if the kernels are very narrow it doesn't much matter, unless there
are large distant sidelobes.  Sometimes authors use merely the locations 
where they tried to centre the kernels, rather than where they
actually succeeded in centring them, which requires conjecture on the
part of the reader relating to the information contained
in the data set employed and the proficiency of the author in
extracting it.  One useful  rule of thumb for helping to guess what
might have been plotted is to recognize that with currently available 
frequency data well localized kernels
without substantial sidelobes cannot be centred closer to the centre
of the Sun than $r/R \approx 0.05$.

Inversion is not a well prescribed procedure.  It is an art.  And it can lead to
different representations of the information, as is illustrated in
Figure \ref{fig4}, which depicts different optimally localized sound-speed
averages all derived from the same data. It should therefore be
recognized that the differences between the averages plotted in the
figure are not necessarily an indication of 
inversion error;  instead, they can result from the differences in the
averaging kernels selected.  It is evident, therefore, that merely
offering some measure of the widths of the kernels, which is absent from Figure \ref{fig4}, is insufficient for 
appreciating the results fully.  

The objectives of most of our investigations do not concern seismic
variables alone.  I have already mentioned issues concerning the solar
neutrino flux, and also the helium abundance [$Y$] and the relation
between its initial value [$Y_0$] and Big-Bang nucleosynthesis.
Investigation of these non-seismic quantities using seismology
necessarily involves relating them to seismic variables via a
theoretical model, which itself depends on the assumptions upon which
that model has been built.  That statement may seem obvious.  But those not in our subject do not
always appreciate  just which of those
assumptions are important.  It therefore does no harm to state them.
Evidently, which of the assumptions  are the most important depends on the matter in hand.
 
I conclude this part of my discussion with another simple point, which
is also not obvious to everybody: the dominant physics of helioseismology is
extremely simple;  it is simply the physics of the propagation and
interference of well understood waves.  Therefore, provided one is
scrupulous in presenting the results with due care and attention, the
direct conclusions cannot be questioned.  When an observation appears
to be in conflict with our seismological knowledge, as superficially
seems to be the case with modern spectroscopic determinations of the 
photospheric heavy-element abundances, for example -- an issue to which
I shall turn my attention below -- contrary to the opinion of some commentators 
\citep[\textit{e.g.}][]{guziketalrevisedabundances2006MmSAI.77.389G} it is not the seismology itself that is to be challenged\footnote{I am assuming that the seismic data processers have judged their errors correctly, and that the error correlations in the frequency data sets, normally ignored,  are not unduly severe.  I am also assuming that, where it is appropriate, asphericity of equilibrium structure is taken correctly into account.}.   I hasten to add that there certainly are issues within
seismology that have not been resolved, usually because a suitable way
to analyse the data has not yet been found (possibly because the data
do not even contain that information in a form that permits it to be readily extracted, possibly
because it is impossible to do so, possibly because the extraction procedure itself still seems to 
be beyond our capabilities), but that is a different matter.

One further matter concerns the manner in which we use a diagnostic.
That typically depends on the issue one wishes to address, and its
relation to that issue may itself be subject to some doubt, perhaps
due to untested assumptions in modelling.  This means that the
accuracy of the result may be much less than the precision, perhaps very much less.  It is
important to recognize the difference between the two.  Precision can
usually be estimated well from the precision of the data -- although in order to trust the outcome 
one has to trust the estimated errors in the data.  Accuracy requires recognizing and
assessing the influence of the assumptions -- inaccurate assumptions can lead
to a dispersion amongst the outcomes in any investigation that prudently utilizes a
variety of different analysis techniques \citep[\textit{cf.}][]{DOG2012Fujihara}.  Therefore precision
is never judged to be lesser when fewer analysis procedures are 
considered.  Unfortunately,  greater precision is often mistaken for greater accuracy.

\begin{figure*}
\centering
     \begin{center}
       \includegraphics[width=8.0cm]{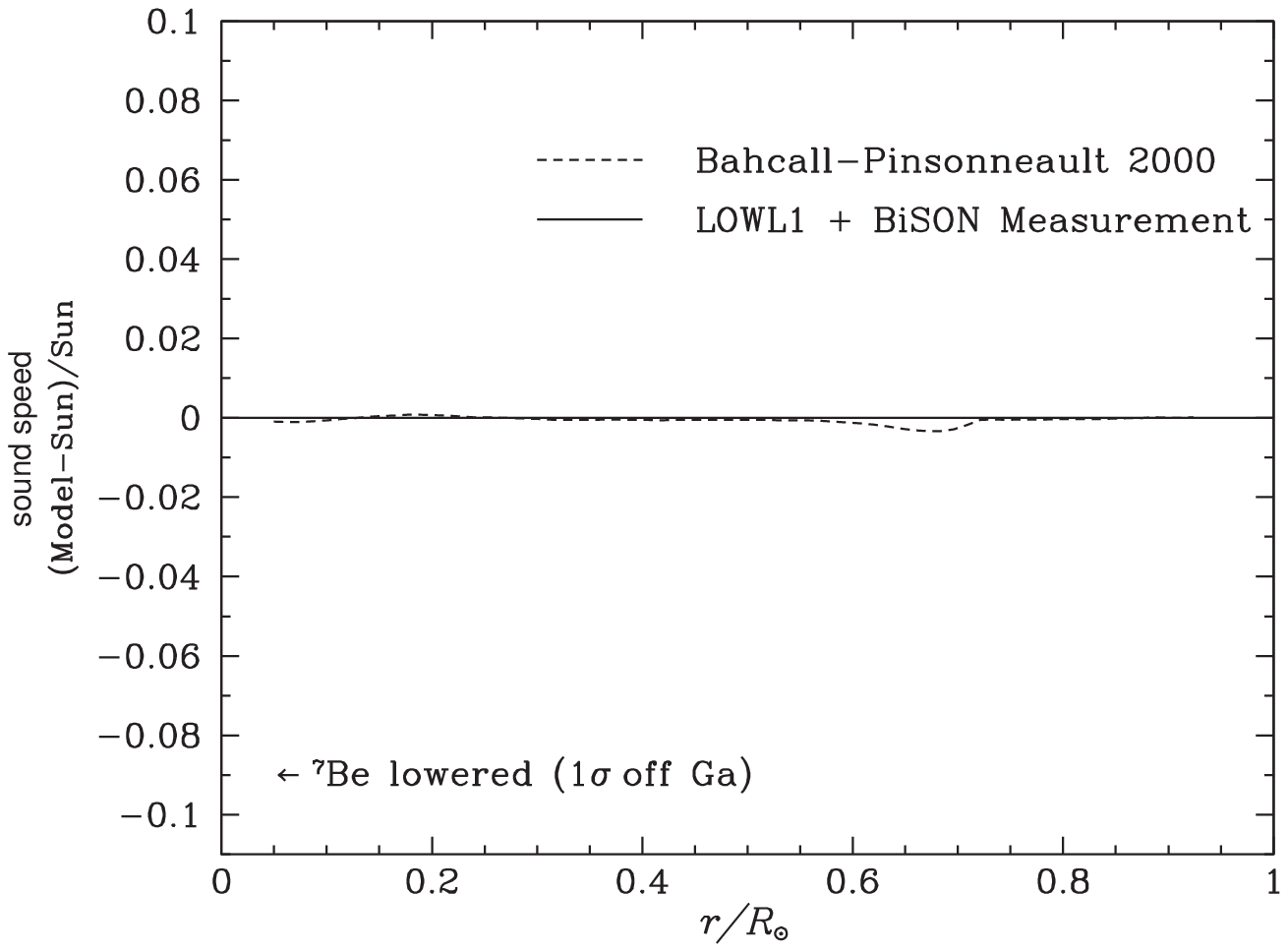}
     \end{center}
\caption{The dashed line is the relative difference between the sound speed in a standard solar model computed by Bahcall and Pinsonneault and that inferred by Basu \citep{Basuetalbestfrequencies1997MNRAS} from a combination of oscillation frequency data obtained by \citet{chaplinetalfrequencies1996SoPh.168.1C} and \citet{tomczykschoumjt1995ApJ...448L..57T,tomczyketalLOWL1995SoPh..159....1T}   \citep[from][]{Bahcallpinsonneaultbasu2001ApJ}.  The arrow labelled $^7$Be \textit{etc.} is simply a very rough estimate of the sound-speed discrepancy in a theoretical solar model adjusted in a manner described by \citet{bahcallbasupinsonneault1998PhLB} to yield a neutrino flux concomitant with that measured by the gallium detector at Gran Sasso \citep{GallexIII1996PhLB.388.384H}.}\label{fig5}
\end{figure*}

\section{Standard Solar Models} 
Standard solar models are constructed usually with the most sophisticated microphysics available, 
yet with the most primitive macrophysics -- fluid dynamical processes are ignored wherever possible, 
probably because they are too difficult to model in an agreed standardized manner.  Rotation and magnetic fields are normally ignored too, so the star is spherically symmetrical.  The star is evolved hydrostatically either from somewhere on the Hayashi track, gravitationally contracting 
until it reaches the main sequence, whence radiative energy loss is balanced almost exactly by nuclear energy generated in the core;  or it is evolved from an estimate of the zero-age main-sequence structure, which is obtained by equating radiant energy loss with nuclear generation in a completely homogeneous star.  The second approach is not quite correct, because there was some nuclear transmutation prior to arrival on the main sequence, most notably conversion of $^3{\rm He}$ to $^4{\rm He}$, which temporarily slows down the contraction (but is insufficient to halt it); however,  that phase has little impact on the subsequent evolution of the star.  Although the transition from gravitational to nuclear energy release is smooth, during subsequent evolution on the main sequence the central hydrogen abundance [$X_{\rm c}$] declines almost linearly with time \citep{DOGlinearXc-t1995ASPC}, so backwards extrapolation of $X_{\rm c}$ to the initial hydrogen abundance [$X_0$] provides a useful fiducial origin of main-sequence age.

\begin{tabular}{lccccccccc}
\noalign{\smallskip}
\noalign{\smallskip}
\hline
\noalign{\smallskip}
\hfil Table 1:   Standard model assumptions\hfil\\
\noalign{\smallskip}
\hline
\noalign{\smallskip}
Initially uniform chemical composition\\
Spherical symmetry\\
Hydrostatic equilibrium\\
Simple description of energy transport, using mixing-length theory in convection\\
~~~~~~zones, and a simplified treatment of radiative transfer in the atmosphere\\
No serious internal mixing of chemical species except in convection zones,\\
~~~~~~so no tachocline, yet gravitational settling and chemical diffusion\\
Rotation dynamically negligible\\
Maxwell stresses negligible (\textit{i.e.} no magnetism)\\
No mass loss\\
No accretion\\
No large-scale instability\\
No (nonlinear) transport by waves\\
Thermal balance (almost)\\
Equation of state known\\
Nuclear reaction rates (reaction cross-sections) known\\
Opacity formula known\\
\noalign{\smallskip}
\hline
\end{tabular}

Evolution on the main sequence is hydrostatic, and essentially in thermal balance, energy being transported from centre to surface by optically dense radiative transfer wherever the stratification is convectively stable: I present the transfer equation in the next section.  Nuclei undergoing thermonuclear reactions are usually considered to be screened by electrons according to the classical, non-relativistic, Debye--H\" uckel treatment by \citet{salpeterscreening1954AuJPh.7.373S}, although the validity of that has been questioned, most recently by \citet{katiewerner2010Ap&SS,katiewerner2011ApJ}.  In convectively unstable regions, the transport of heat is modelled by a (usually local) mixing-length formalism; Reynolds stresses are usually, but not always,  ignored.  Chemical species are presumed to be homogenized in convection zones; and elsewhere, aside from nuclear transmutation,  chemical differentiation is purely by gravitational settling (and possibly radiative levitation) against microscopic diffusion.  Energy and material transport by the tachocline circulation is normally ignored, as is transport by acoustic and gravity waves generated by the turbulent convection.  The total mass of the Sun is presumed to be conserved, and the ``best'' equation of state, opacity ``formula'' and nuclear reaction cross-sections are employed.

It goes without saying that in computing  hydrostatic structure it is preferable (some would say mandatory) 
that the coordinate singularity at the centre of the star and the effective  branch point  at the radiative-convective interface (when a local mixing-length formalism is adopted) be correctly treated by the numerical integration, accurately enough for the resulting solar model to possess well defined seismic eigenfrequencies.  Unfortunately, that is not always achieved, although I suspect that the numerical precision is greater than the observational accuracy.

\section{The Chemical Composition of the Sun} 
Some time ago solar modelling \citep[\textit{e.g.}][]{jcdetal1996Sci}
had achieved fair agreement with the measured seismic structure of the
Sun \citep[\textit{e.g.}][]{Goughetal1996Sci}.  Indeed, \citet{bahcalltriumphforstellarevolution2001Natur} considered
that to be a ``triumph for the theory of stellar evolution'', which he
illustrated by plotting the relative deviation of the solar sound
speed from that of one of his standard solar models \citep[\textit{e.g.}][]{Bahcallpinsonneaultbasu2001ApJ}.  It is
reproduced here as Figure \ref{fig5}.  Yet might one not naively consider it to
be an even greater triumph for helioseismology?   After all,  the
seismological errors are much smaller than the modelling errors, as is
more clearly evident in Figures \ref{fig3} and \ref{fig4}.  But I have already
emphasized that the physics of seismology is very simple, whereas the
physics of the structure of the Sun is not.  The comparison can
therefore be regarded in two opposite ways:
i) one might marvel, as did Bahcall, that such complicated physics has
been successfully reined to reproduce observation (more or less), or
ii) one might consider it unremarkable that a mature theory, which has
already survived a wide range of astronomical tests relating to other
stars, and which depends on so many somewhat uncertain processes, can
be adjusted to reproduce the measurable properties of the Sun.  I
think I side with Bahcall on this matter, because nobody has actually
succeeded in reducing the discrepancy between the models and the Sun
sufficiently to come within the much smaller seismological uncertainty.  Moreover,
the situation has been exacerbated by the results of recent
spectroscopic re-analyses,  by \citet{Asplundetal2009ARA&A}, \citet{ caffauetal2009A&A, Caffauetal2011SolPh} and \citet{GrevesseAsplundSauval2011sswh.book},  of some
of the chemical abundances in the solar atmosphere, those analyses now
taking explicit account of spatial inhomogeneity in the Sun's atmosphere caused by convective
motion as simulated by  \citet{asplundetal2000A&A...359..743A}  based on a compressible hydrodynamical procedure described by \citet{steinnordlundatmosprops1998ApJ}\footnote{The original analysis employed simulations in a model atmosphere with the previously accepted chemical composition; from 2009 onwards, newer simulations were carried out with the Fe abundance inferred from the 2005 abundance determinations; the abundances of C, N, and O were not changed because their spectral lines are weak and have no significant impact on the radiative energy flux (R. Trampedach, personal communication, 2012).}.  The early publications \citep{AllendePrietoetalsolarabundances2001ApJ, AllendePrietoetalsolarabundances2002ApJ, asplundetal2004A&A, Asplundetal2005, asplundetalerratumof20042005A&A, asplundetal2005A&A}   suggested an enormous reduction in 
the total heavy-element abundance $Z$ below previously accepted values:  some 40\,\% or so.  If that represents the abundance in the radiative interior, it implies, on average, a reduction in opacity of some 30\,\%.  By present-day standards, that figure is enormous. 

One of the best and most widely adopted solar models (combining all
the generally accepted physics for so-called ``standard'' theory into a
model computed accurately enough for reliable computations of its oscillation
eigenfrequencies to be possible) is the Model S discussed by
\citet{jcdetal1996Sci}.  It was computed with a local mixing-length theory ignoring turbulent pressure, with mixing-length parameter [$\alpha$] and 
initial
hydrogen and heavy-element abundances 
[$X_0 = 0.7091,\; Z_0 = 0.0196$ ]
chosen to yield the correct present-day radius and what was believed to be the correct luminosity,  and also a resulting present surface abundance ratio [$Z_{\rm s}/X_{\rm s}$] of  
0.0245,  consistent with the spectroscopic analysis of \citet{Grevessenoels1993}, at an age $t_\odot = 4.60$ Gy (notwithstanding the different value quoted  by  \citet{jcdetal1996Sci}, and subsequently by \citet{doganboannoJCDsurfaceeffectsandage2010}).  The individual surface abundances
are $X_{\rm s}=0.7373,\; Z_{\rm s}=0.0181$; the present surface helium abundance is
$Y_{\rm s}=0.2447$, which agrees with observation within the limits of accuracy of the equation of state and 
helioseismological analyses of the depression of $\gamma_1$ caused by
the second ionization of helium in the adiabatically stratified
convection zone \citep{DOGCatania1984MmSAI, agketalsolarY1992MNRAS, svvvalodiaaloshaeos1992MNRAS,  basuantiasolarY1995MNRAS.276.1402B, werner2007AIPC..948..179D}, as I shall discuss briefly later.  Figures \ref{fig3} and \ref{fig4} illustrate how 
well the seismic structure of the model corresponds to that of the Sun.

The heavy-element abundance now favoured by \citet{GrevesseAsplundSauval2011sswh.book} is
somewhat greater than the original announcement, namely 
$Z_{\rm s} = 0.0134\,\pm \, 0.0005$; an independent and somewhat different analysis by \citet{Caffauetal2011SolPh} that also takes convective inhomogeneity into account has yielded 
$Z_{\rm s} = 0.0153\,\pm\,0.0011$.  Both of these values were estimated from
abundances of the major opacity-producing elements except
neon, whose abundance in the Sun cannot be measured
spectroscopically.   A recent compilation by \citet{lodderspalmegail_solarZ_2009}
using a wider range of solar data, with an eye also on meteoritic abundances, yet using simple hydrostatics, has yielded the  recommendation $Z_{\rm s}/X_{\rm s}=0.0191$.  Accepting 
$Y_{\rm s} = 0.249$ from a calibration of model envelopes by \citet{basuantiasolarabundances2004ApJ.606L.85B} using seismically accessible (asymptotic) integrals that are sensitive to the depression of $\gamma_1$ due to He\,{\sc ii} ionization, or  $Y_{\rm s} = 0.224$ from a calibration of complete solar models  \citep{GHDOGsolarage2011MNRAS}, yields $Z_{\rm s}=0.0141$ and $Z_{\rm s}=0.0145$, respectively.  (It is unclear whether calibrating complete solar models in this fashion is more reliable or less reliable that calibrating only model envelopes.)
These values are all substantially lower than that of Model S, on average by about 20\,\%.
What are the implications of these results?

\subsection{The Abundances in Context}
Many of the discussions that have ensued have carried out
solar-evolution computations and then compared, in some manner, 
implied seismic data, or seismic structure (principally sound speed),
with the inferences from the Sun.  They have been catalogued by \citet{basuantiaabundancereview2008PhR}.  However, seismic structure is merely a diagnostic
of the issue, not the issue itself.  It is instead much more lucid to the
average physicist to be confronted with the direct implication of the
newly reported abundances.  So let us accept (most of) the assumptions
of solar modelling, as is usually done in the discussions of the
problem, and ask what they really imply.  I start with the most obvious: that
the Sun approached the main sequence fully mixed, having just come
down the Hayashi track.  Therefore the heavy-element abundance [$Z(r,t_\odot)$]
in the radiative interior today is (almost) the same as it is in the photosphere,
aside from a relative excess of 3\,\% or so due to differential
gravitational settling.  Heavy elements influence the structure of the
Sun primarily via their effect on the opacity [$\kappa$], which controls the
relation between luminosity [$L$] and temperature [$T$] through the
equation of radiative transfer: $L(r,t) = - \left (16 \pi ac \,r^2
T^3/3 \kappa \rho \right ) {\rm d}T/{\rm d}r$.  However, $T$ is not a seismic variable,
and therefore cannot be measured directly.  It is related to the
seismic variables $c^2$ and $\rho$ by the equation of state:
$T={\cal{T}} \left (c^2,\rho;  Y, Z \right)$ -- I set aside, for the
moment, consideration of the relative abundances of the heavy
elements. The contribution of $Z$ to the equation of state is only
about 0.5\,\%, so for the purposes of this discussion its uncertainty can safely be ignored.  There remains
only the helium abundance [$Y$] which is to be estimated from the
theory of stellar evolution.  It is important to appreciate that this
is the only stage in the argument where the details of the evolution
theory come into serious play.   However, it must be realized that knowledge of  
the functional form of 
$Y(r)$ today is crucial for assessing the stratification of the Sun's radiative interior.  Therefore, some 
attention must be paid to how it is determined, appreciating the assumptions to which the outcome is sensitive.

As the Sun evolves on the main sequence, hydrogen is transmuted into
helium in the core, adding a spatially and temporally varying component $\delta_\varepsilon Y$
to the helium abundance:  $Y(r,t) = \left[1+s_Y(r,t) \right] Y_0 +\delta_\varepsilon Y$, where
$s_YY_0$ is the change in $Y$ produced by gravitational settling
moderated by diffusion.  It is known \citep[\textit{e.g.}][]{DOGprotosolarY1983,1990stromgren} that the time
dependence of the total luminosity $L_{\rm s}(t):=L(R,t) $ of the solar
model, calibrated to satisfy $L_{\rm s}(t_\odot) = L_\odot$, is not very sensitive 
to the details of the theory (such as the choice of $Z_0$, or whether
or not there has been some small degree of material mixing in the core -- we know from
seismology that the core cannot have been homogenized, a conclusion
which is consistent with HR diagrams of solar-like stars), so the
total amount of helium that has been produced by 
today [$\int \delta_\varepsilon Y{\rm  d}r $] is proportional to $\int L_{\rm s}{\rm d}t$, which is essentially known (provided $t_\odot$ is known).
Therefore it is adequate for the current discussion to accept the function 
$ \delta_\varepsilon Y$ from any
standard model.  Also, the function $s_Y(r,t)$ is only very weakly
dependent on $Y_0$, and may also safely be taken as given by the model.   
Likewise, one can take $Z= \left[1+s_Z(r,t) \right ] Z_0$ with $s_Z$ 
given by the model  (it is broadly similar to the function $s_Y$). 
Whence $T(r,t)={\cal{T}} \left(c^2, \rho;\left(1 + s_Y \right) Y_0 + 
\delta_\varepsilon Y, \left(1+s_Z \right) Z_0 \right)$, 
where $c^2$ and $\rho$ have the values determined seismologically from the Sun.  
Only the initial abundance $Y_0$ now remains unknown.  How do 
we determine it?

\begin{figure*}
\centering
     \begin{center}
       \includegraphics[width=9.0cm]{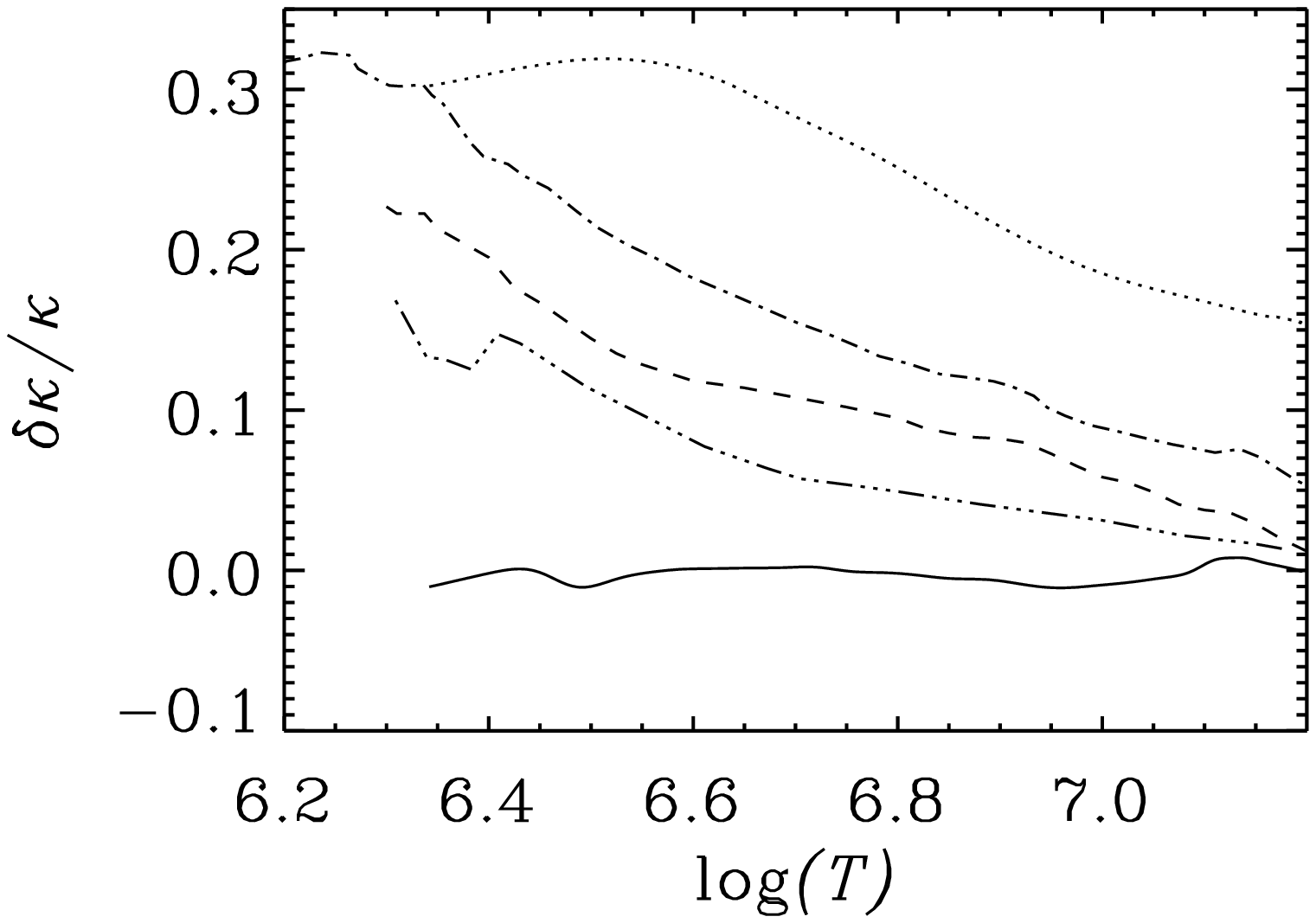}
     \end{center}
\caption{The dot--dashed curve is the estimate by \citet{JCDetalopacity2009} of the relative opacity augmentation required when adopting  the
\citet{asplund2005ARA&A.43.481A} chemical composition in OPAL opacity computations \citep{updatedopacity1996ApJ.464.943I} to produce a solar model with the same sound speed as in Model S  of \citet{jcdetal1996Sci}.   The dashed curve represents  the augmentation, computed  by \citet{jcdhgreview2010Ap&SS.328...51C}, required of OPAL  opacities  with the revised \citet{Asplundetal2009ARA&A} abundances.  The triple-dot--dashed curve is a comparable difference between two solar models computed by \citet{jnbsbmhpamsnewabundances2005ApJ...618.1049B} with OPAL opacities, the first with the latest abundances by Asplund \textit{et al.} at the time, the second with abundances that have been artificially chosen to yield what one might regard as a seismically acceptable proxy Sun.   For comparison, the dotted curve is a linearized estimate of the same quantity were the relative abundances of the heavy elements to have been preserved, using opacity derivatives obtained from the tables of \citet{jnbmhpstandardmodels1992RvMP...64..885B}. The continuous  curve is the relative difference $(\kappa_\odot - \kappa_{\rm S})/ \kappa_{\rm S}$   inferred by \citet{DOGLorentz2004AIPC} between the Sun's opacity [$\kappa_\odot$]  and that of Model S.  }\label{fig6}
\end{figure*}

I recommend calibrating $ Y_0$ by accepting the nuclear reaction rates -- 
after all, all the pertinent nuclear cross-sections have been carefully assessed 
experimentally during the investigations of the neutrino problem, save 
the controlling p--p reaction cross-section which is determined only 
theoretically (however, it is the simplest of all nuclear reactions). 
There remains also an issue concerning the screening of energetic 
particles \citep[\textit{e.g.}][]{katiewerner2010Ap&SS, katiewerner2011ApJ}.  These matters are 
probably, for the purposes of this discussion, minor.  Therefore the
total rate of thermal energy production, hence the luminosity, is
determined in terms of $t_\odot$ and $ Y_0$.  Granted that
$L_\odot$ and $t_\odot$ are known quite well, $Y_0$ can thus be
calibrated to determine $T(r)$ today\footnote{Implications from earlier estimates of $Y$ to determine $T(r)$ have been discussed by \citet{elliottopacity1995MNRAS.277.1567E} and \citet{tripahyjcdI_1998A&A...337..579T}, the first under the assumption  that $Y(r)$ differs from that in a reference solar model by just a constant,  the second that it can be obtained simply by scaling the reference-model value by a constant factor, both of which are inaccurate in the core, although Elliott notes that the opacity perturbations produce a predominantly  local response, so that these analyses should provide a fair estimate in the radiative envelope.
The Tripathy--Christensen-Dalsgaard scaling was used by \citet{tripathybasujcdopacity98} to obtain the opacity difference from Model S by RLS frequency fitting (J. Christensen-Dalsgaard, personal communication, 2012)  assuming that that difference can be expressed as a function of $T$ alone, yielding a superficially similar functional form to the continuous curve in Figure 6, but with a magnitude about 50 per cent greater.   \citet{jnbsbmhpamsnewabundances2005ApJ...618.1049B} have estimated the opacity difference by adjusting 
$\kappa$ by hand in solar models.   Their preferred model had a constant 11\,\% augmentation over the OPAL values using the most recent abundance determinations 
\citep[][Asplund  personal communication with Bahcall \textit{et al.}, 2004]{asplundetal2000A&A...359..743A,asplundetal2004A&A,  asplund2005ARA&A.43.481A, AllendePrietoetalsolarabundances2001ApJ, AllendePrietoetalsolarabundances2002ApJ} 
in the radiative envelope down to $T=5\times 10^6$K, beneath which the augmentation was smoothly reduced to zero (probably by a half Lorentzian function with half-width at half maximum of $2\times 10^5$K), and has a seismic structure as close to that of the Sun as does Model S.  
Regarding that model as a proxy Sun, one would expect the relative opacity difference between it and a model with the unmodified abundances to be comparable with the inference by \citet{JCDetalopacity2009}.  The two estimates are depicted in Figure 6.  
\citet{korzennikulrichopacity1989ApJ...339.1144K} had earlier estimated opacity errors by RLS (${\rm{L_2}}$ norm) data fitting, and \citet{saio_opacity1992MNRAS.258..491S} by ${\rm{L_1}}$   data fitting, each by scaling $\kappa$ by a function of $T$ and ignoring the dependence of the relation between $T$ and the seismic variables on chemical composition; they expressed their results as deviations from different reference models, so they cannot easily be compared with those presented in Figure 6.  
An estimate by OLA \citep{takatgough2001ESASP.464..543T} of the absolute structure of the Sun, including opacity, using the procedure for determining $Y$ described in the text (together with tachocline homogenization as calibrated by 
\citet{elliottdogsekiitachocline1998ESASP}) is presented by \citet{dogphs2002css1.book.1035G} and \citet{DOGLorentz2004AIPC,DOGNaxosconf2006ESASP.617E.1G}.}, 
which can then be substituted into the
equation of radiative transfer to evaluate $\kappa(r)$.  
That procedure can either be carried out explicitly from a seismological inversion to determine 
the seismic stratification, or implicitly within the inversion itself, such as in the manner advocated 
by \citet{elliottopacity1995MNRAS.277.1567E}.  The
outcome is illustrated in Figure \ref{fig6}, in which the continuous curve is the relative difference 
$(\kappa_\odot - \kappa_{\rm S})/ \kappa_{\rm S}$ between the Sun's opacity [$\kappa_\odot$] determined in the manner that I have just described 
and that of Christensen-Dalsgaard's Model S.
Not surprisingly, $\kappa_\odot$ 
is very similar to the opacity in Model S.
The issue posed by Asplund \textit{et al.}  and Caffau \textit{et al.} is therefore simply an abundance--opacity problem: 
how  can $\kappa_\odot(r)$ be reconciled with their abundance measurements?

\subsection{Suggestions for Reconciliation}
An obvious naive suggestion is that
opacity calculations are in error by just the appropriate  factor
required to compensate for the proposed revision in $Z$, leaving the
functional dependence on the other variables  unscathed. Given the testing that has been undertaken by
Iglesias and Rogers, that does not seem so very likely.  It is also
unlikely that energy transport is opposed by some other mechanism,
such as gravity-wave transport, because that would require a
completely different physical process to mimic the functional form of the effect of 
radiatively induced atomic transitions.  Of course, it could be that
the new abundance determinations are in error, despite the extra care
that has been taken; we have seen an increase since the early work by 
 \citet{AllendePrietoetalsolarabundances2001ApJ, AllendePrietoetalsolarabundances2002ApJ} and  \citet{ asplundetal2004A&A,Asplundetal2005, asplundetalerratumof20042005A&A, asplundetal2005A&A}, so a further increase
might not be too surprising.  However,  the independent investigation
by  \citet{ caffauetal2009A&A, Caffauetal2011SolPh} has yielded results not very dissimilar  from those of \citet{Asplundetal2009ARA&A} and \citet{GrevesseAsplundSauval2011sswh.book}, 
which adds to the credibility of the new abundances. 
Of course, another possibility is that early in its post-Hayashi days the
Sun was contaminated by metal-poor material, as \citet{guziketalsolarcontamination2005ApJ} have suggested.  If that were the case, it may just be possible to detect seismologically a relic compositional discontinuity at the location of the base of the zero-age convection zone, 
provided that it has not been destroyed.   
The discontinuity is at least stable to double-diffusive convection, so it would not be destroyed spontaneously.  Yet a fourth possibility is that the
relative abundances of the heavy elements are different from what is
generally believed, particularly that of neon, as has been discussed,
for example, by \citet{draketesta2005Natur}, leaving open the possibility that
$Z$ is greater than the values suggested by Asplund \textit{et al.} and Caffau
\textit{et al.} based on the assumption that the relation of the neon abundance  to the abundances of the 
other opacity-producing elements is preserved, and that therefore $\kappa$ is perhaps proportionately greater too.  That too is regarded by many spectral analysts as being intrinsically unlikely, and in any case neon alone does not mimic the heavy-element mixture adequately to restore the solar models to their pre-Asplund state.

The apparent opacity discrepancy has evolved with the abundance revisions.  
\citet{JCDetalopacity2009} estimated the amount $\delta \kappa$ by which the opacity according 
to OPAL \citep{updatedopacity1996ApJ.464.943I} with \citet{asplund2005ARA&A.43.481A} chemical composition would need to be augmented in order to produce an otherwise standard solar model with the same sound speed as in Model S, and hence about the same sound speed as in the Sun. 
It is depicted as the dot-dashed curve in Figure \ref{fig6}, where it is plotted against log($T$) along the thermodynamic $\rho$--$T$ path of one of the models.  One might have expected the outcome to have been given approximately by $(\partial~ {\rm ln}~\kappa/\partial~ {\rm ln}~Z)_{\rho,T,X}~\delta~ {\rm ln}~Z$, in which $\delta ~{\rm ln}~Z \approx {\rm ln}~Z_{\rm S}-{\rm ln}~Z_{\rm Asp}\approx {\rm ln}~Z_{\rm s,S}-{\rm ln}~Z_{\rm s,Asp} = 0.370$, the subscript s denoting surface value and the subscripts S and Asp denoting Model S \citep{Grevessenoels1993} and \citet{asplund2005ARA&A.43.481A} values respectively, and where the opacity derivative is evaluated at constant relative heavy-element abundances.  However, it is evident from Figure \ref{fig6} that that is not the case.  According to J\o rgen Christensen-Dalsgaard (personal communication, 2011) the quite substantial difference arises because the relative heavy-element abundances in the two models differ.  
Unfortunately, that renders back-of-the-envelope estimates suspect. 
The later abundance determinations by \citet{Asplundetal2009ARA&A} have brought the OPAL opacities closer to those of the Sun, as the dashed curve in Figure 6 depicts.  
It is interesting that  \citet{antiabasuabundancereconciliation2011JPhCS}, using the chemical composition  proposed by \citet{Caffauetal2011SolPh}, whose relative abundances do differ from the earlier values given by \citet{grevessesauvalabundances1998SSRv}, and \citet{Grevessenoels1993}, and presumably also the more modern values of \citet{GrevesseAsplundSauval2011sswh.book}, report that they have obtained solar models that are seismically  almost as good as models constructed with the abundances of \citet{grevessesauvalabundances1998SSRv}, and they found even closer agreement with the Sun if the
assumed abundance of neon were artificially enhanced by a factor $\sqrt{2}$.  That essentially reduces the problem of reproducing 
the Sun's seismic stratification theoretically to the state in which it was prior to Asplund's original announcement.  It must be realized, however, that that does not close the matter,  because merely reproducing previous partial results does not necessarily prove the veracity of the physics behind 
the new models.  Moreover, the seismic structures of the models, new and old,  deviate from that of the Sun by many standard errors, as is evinced by Figure \ref{fig3}.

\subsection{Possible Flaws in the Argument}
The foregoing discussion is predicated on the presumption that the sole significantly discrepant 
ingredient in the solar modelling is opacity.  That need not be the case; modification of any 
non-seismic variable might, at least in principle, bring a theoretical model into line with observation. Evidently,  
a  direct change in  the relation between the seismic variables pressure and density, for example,    
via a modification to the abundances of the abundant chemical elements H and He, can alter the sound speed:  \citet{DOGAGK1988ESASP.286..195G, DOGAGK1990ASSL..159..327G} found that 
the stratification of the energy-generating core can be reproduced by a slight smoothing of the abundance profile that had been produced by nuclear transmutation, suggesting that a slight 
degree of mixing has taken place, possibly by disturbances that have been shear-generated in a manner analogous to clear-air 
turbulence in the Earth's atmosphere.  Alternatively, there could be an additional source of energy transport without mixing, such as by accreted weakly interacting massive particles (wimps), whose presence in the Sun was postulated originally in order to try to account for the observed low neutrino flux without neutrino transitions \citep{spergelpresswimps1985ApJ...294..663S}; wimps   
modify the temperature distribution, and consequently the distribution of helium produced 
by nuclear transmutation, leading to a sound-speed modification whose general functional form is broadly similar to what would 
be required \citep{gfps_wimpysolarmodels1986ApJ...306..703G}, although a serious attempt to reproduce the seismic structure of the radiative interior appears not to have been made \citep[see also][]{faulknergoughvahia1986Natur.321..226F, wdrlgjcdwimps1986Natur.321..229D, jcdwimps1992ApJ...385..354C}.

It is worth commenting that a reduction in $Z_{\rm s}$ at fixed relative abundances of only about the 20\,\% below that of Model S  implies, via
the theory of solar evolution, a reduction of about 0.02 in $Y_{\rm s}$.
This appears to me to be not dissimilar to the margins of the true uncertainty in the
helioseismological determinations of the helium abundance of the convection zone -- although not within the precision
quoted by those who have attempted to determine $Y_{\rm s}$ by only a single procedure 
\citep[\textit{e.g.}][]{basuantiasolarY1995MNRAS.276.1402B,basuantiasolarabundances2004ApJ.606L.85B,basu1998MNRAS.298..719B,RichardDziembowskietalprecisesolarY1998ESASP.418.517R,
dimaurobasuetalintrinsicgamma_12002A&A}, even granted that systematic errors had been recognized.

\subsection{Seismological Investigation}
Chemical abundances can in principle be measured directly by seismology by analysing the non-ideal properties of the solar plasma.   Specifically, $\gamma_1$ is depressed in regions of partial ionization, by an amount that is almost proportional to the abundance of the ionizing element. 
\citet{WDDOGsolarYLiege1984LIACo.25.264D} proposed measuring the helium abundance by calibrating the ionization-induced variation of a thermodynamic function [$\Theta$] expressible in terms of $\gamma_1$ and its derivatives and which is easily  accessible to seismological probing in the adiabatically stratified layers of the convection zone \citep{DOGCatania1984MmSAI}.  A procedure for so doing was developed by \citet{wddogmjtfurtherprogressforY_1988ESASP.286..505D}, who found $Y_{\rm s}$ to be lower than typical values obtained from calibrating evolved solar models \citep{WDDOGMJTChallenges1991LNP...388..111D}, which triggered \citet{JCDProffittMJT1993ApJ.403L.75C} to investigate the seismological implications of gravitational settling of heavy elements, with notable success.  Other forms of calibration, often based more directly on $\gamma_1$, have been pursued \citep[\textit{e.g.}][]{WDDOGMJTChallenges1991LNP...388..111D, basuantiasolarY1995MNRAS.276.1402B,basuantiasolarabundances2004ApJ.606L.85B,basu1998MNRAS.298..719B,RichardDziembowskietalprecisesolarY1998ESASP.418.517R,
dimaurobasuetalintrinsicgamma_12002A&A, HGseconddiff2007MNRAS}.  None has been completely satisfactory, not least because there are serious uncertainties in the equation of state. 
There have been attempts to estimate $Z$ by calibrating solar models against low-degree frequency-separation ratios assuming $t_\odot$ to be known \citep[\textit{e.g.}][]{chaplinetalmodefrequencies2007ApJ...670..872C}; that is equivalent to calibrating $Y$, of course.

A direct seismological estimate of $Z$ in the convection zone,  and thus a direct estimate
of $Z_{\rm s}$, via the ionization-induced depressions in $\gamma_1$  or in the variation of the 
seismologically more accessible function $\Theta $ of $\gamma_1$ and its derivatives, is very delicate.  It would probably be
necessary to average the individual variations in the different ionization
zones, so the result would depend to some degree on the assumed values
of the relative abundances \citep{kmdog2009}.  Nevertheless,
if successful, the accuracy of the result could probably be assessed
from the seismic frequency uncertainties, because the relation between
the magnitude of the $\gamma_1$ depression and the abundance of any 
ionizing element producing it, at least at the level required to judge the abundance--opacity problem, is fairly robust.  The seismologically apparently more
straightforward procedure of measuring a spatially averaged absolute value of
$\Theta$ and relating that to the slightly greater value expected in the absence of
heavy elements, as \citet{antiabasuZdetermination2006ApJ} have tried, is more
uncertain because the change in $\Theta$ due to heavy-element
ionization that is sought appears to be less than  the uncertainty 
in the equation of state arising, for example, from the commonly neglected, at least in the chemical picture 
\citep{dappenlorentz2004AIPC, werner2007AIPC..948..179D},  finite volume occupied by bound species \citep[\textit{e.g.}][]{BaturinWDDOGSVV2000MNRAS}. 

\subsection{The Equation of State}
My statements about the uncertainty in $Y_{\rm s}$ and the equation of state call for some justification.  It goes without saying that chemical composition is not a seismic variable, not even the abundances $X$ and $Y$ of the abundant elements H and He.  In order to determine $Y$, say, it is evidently necessary to use an equation of state, which relates the seismic variables to composition via another, necessarily non-seismic, variable, such as $T$ or $s$, and subsequently, at some point, to confine attention to the adiabatically stratified region of the convection zone where that variable   can be eliminated from the spatial variation of the seismic variables. Once that has been accomplished one can learn about chemical composition through the effect of ionization on $\gamma_1$.  This has been approached either directly by measuring some appropriate property of  $\gamma_1$ itself, or by working with some thermodynamic function $\Theta$ of it.  The reliability of the outcome necessarily rests on the reliability of the equation of state, which is very difficult to assess.  That should be obvious because there is no redundancy in the dependence of the seismic modes on the seismic variables.  Therefore, as has been alluded to in the past \citep[\textit{e.g.}][]{DOGLorentz2004AIPC}, perhaps too obliquely, any intrinsic error in the equation of state cannot be assessed by seismology alone.  To make progress, non-seismic information must be incorporated.  In practice, that information comes from some prior appreciation of the reliability of some aspects of the equation of state, coupled with the additional non-seismic constraint that the convection zone is chemically homogeneous (sometimes augmented with the constraint that deep down the stratification is adiabatic).  Suppose, for example, one is estimating the deviation of the Sun from a reference solar model, represented by  $\delta {\rm ln}\gamma_1$  and $\delta {\rm{ln}}u$, where $u(r)$ is a complementary seismic variable.  From seismology one can relate averages $\overline{\delta {\rm ln}\gamma_1}+\overline{\overline{\delta {\rm {ln}u}}}$  of those deviations to data combinations $d$, where the single and double overbars indicate simply that the averaging kernels are different.  Then, one can write the first average as the sum of 
$\overline{(\partial{\rm ln}\gamma_1/\partial {\rm {ln}}u)_Y\,\delta {\rm {ln}}u}$, which can be incorporated into the second average, of 
$\overline{(\partial{\rm ln}\gamma_1/\partial Y)_u}\,\delta Y$,   
where $\delta Y$ is the difference in $Y$ between the Sun and the reference model, and of a component  $\overline{\delta_{\rm{int}} {\rm ln}\gamma_1}$ resulting from intrinsic error in the equation of state.  One would like to be able to distinguish between those components.  However, that is strictly impossible because they  have the same averaging kernel.  Therefore one cannot unambiguously determine $\delta Y$ from the data [$d$].  What has been attempted in the past is to design kernels such that those defining the overbar and the double overbar are both in some sense small, and then to estimate $\delta Y$ by neglecting $\overline{\delta_{\rm{int}} {\rm ln}\gamma_1}$ and $\overline{\overline{\delta {\rm {ln}u}}}$, a procedure which is evidently not strictly valid; after that, $\overline{\delta_{\rm{int}} {\rm ln}\gamma_1}$ can be computed with presumably different and well localized averaging kernels.  The results of such a procedure have been presented by \citet{basudappennayfonov1999ApJ},  \citet{dimaurobasuetalintrinsicgamma_12002A&A}, and \citet{dappenlorentz2004AIPC}, using both OPAL and MHD, the two
most popular, and probably the best, equations of state available today.  Not only do their inferred intrinsic errors from the two equations differ, as they surely must, but so too do the estimates of the helium abundance [$Y_{\rm s}$] in the Sun's convection zone\footnote{Di Mauro \textit{et al.} quote $Y_{\rm s}=0.2539 \pm 0.0005$ when OPAL is used, $Y_{\rm s}=0.2457 \pm 0.0005$ when MHD is used, the errors being merely formal, representing a precision error that takes no account of the error in the relation between $Y_{\rm s}$ and  $\gamma_1$ in the reference model.  With those values the magnitude of the inferred $\overline{\delta_{\rm{int}} {\rm ln}\gamma_1}$ was found to be the greater for the MHD equation of state beneath the helium ionization zones, and the lesser above $r/R\approx0.97$ in the He\,{\sc ii} ionization zone.  That is perhaps not surprising because MHD is possibly better at taking into account the complicated chemistry that dominates higher up in the solar envelope, whereas the virial expansion used for OPAL is perhaps more reliable where such complications make only a minor non-ideal contribution.   The difference between OPAL and MHD must surely offer some estimate of the total uncertainty.}.  It must be appreciated that any seismological inference about $\delta_{\rm{int}} {\rm ln}\gamma_1$ such as this is  susceptible to potential errors in the inferred values of $Y_{\rm s}$, which are difficult to appreciate because the precise manner in which $\delta Y$ was obtained is unclear.  However, well outside the hydrogen and helium ionization zones $\partial{\rm ln}\gamma_1/\partial Y$ is small,  so errors in $\delta Y$ can hardly contaminate $\overline{\delta_{\rm{int}} {\rm ln}\gamma_1}$ there.  This is a robust property of any realistic equation of state, because under normal stellar conditions $\gamma_1$ is essentially independent of $Y$ where hydrogen and helium are both  fully ionized, errors in the small dependency that does remain (arising from processes such as the  influence of the electron density on the ionization of heavy elements)  being characterized by the quantity $\overline{\delta_{\rm{int}} {\rm ln}\gamma_1}$ itself.  However, the situation is different  within the ionization zones, where 
in general $\overline{\delta_{\rm{int}} {\rm ln}\gamma_1}$ and $\overline{(\partial{\rm ln}\gamma_1/\partial Y)}\,\delta Y$ contribute comparably, and cannot be entirely recognized apart.  I must add, however, that $\overline{\delta {\rm ln}\gamma_1}$ and $\overline{(\partial{\rm ln}\gamma_1/\partial Y)}$ 
are normally functionally different, so $\overline{\delta_{\rm{int}} {\rm ln}\gamma_1}$ cannot vanish everywhere.  

\begin{figure*}
\centering
     \begin{center}
       \includegraphics[width=8.0cm]{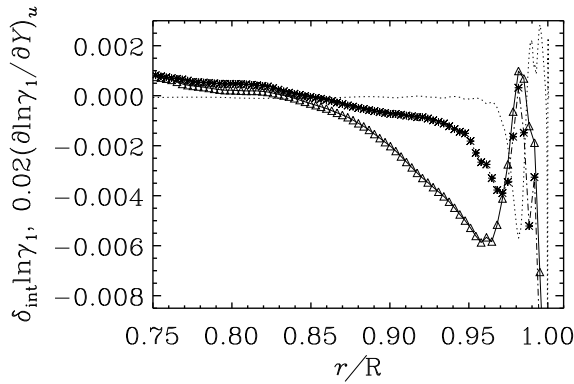}
     \end{center}
\caption{Inferences concerning the intrinsic error in the OPAL (asterisks) and MHD (triangles) equations of state, based on the assumption that the helium abundance, in each case, is such as to minimize its  ${\rm L}_2$ norm 
with respect to acoustic radius over the region of the convection zone that is stratified adiabatically to better than 1 part in $10^3$, computed from the analysis of \citet{dimaurobasuetalintrinsicgamma_12002A&A}.  Because the helium abundance is uncertain, the symbols should not be meant to represent reliable estimates of the actual intrinsic errors $\overline{\delta_{\rm{int}} {\rm ln}\gamma_1}$.   Formal error bars are not drawn because they are enormously smaller than the true uncertainties, and might therefore confuse.  The dotted curve is 
$0.02 (\partial {\rm ln}\gamma_1/\partial Y)_u $, 
computed with the OPAL equation of state.}\label{fig7}
\end{figure*}

An alternative, more transparent,  approach to an attempt at separation could be to adopt the attitude that one's chosen equation of state is as good as it can be, at least in the adiabatically stratified region where $\overline{\delta {\rm ln}\gamma_1}$ has been inferred.  Therefore one could choose for each equation of state the value of $\delta Y$ that minimizes  the integral with respect to acoustic radius of $({\delta_{\rm{int}} {\rm ln}\gamma_1})^2$ over that region -- I choose acoustic radius because that is the natural seismic frequency-controlling independent variable.  The outcome, using the inferences of $\overline{\delta_{\rm{int}} {\rm ln}\gamma_1}$  and $Y_{\rm s}$ by \citet{dimaurobasuetalintrinsicgamma_12002A&A}, is $Y_{\rm s}=0.2430$ for OPAL and  $Y_{\rm s}=0.2280$ for MHD; I do not quote a precision because it is hardly material to estimating  accuracy in this context.  The range of all these values, roughly 0.02, must surely be regarded as a lower bound to the range of values within which the solar $Y_{\rm s}$ is likely to lie, because the discrepancies are systematic; that value is some  40--100 times greater than typical errors quoted by, for example,  \citet{basu1998MNRAS.298..719B} or \citet{dimaurobasuetalintrinsicgamma_12002A&A}, based on the precision of the specific procedures that were carried out.  Based solely on the calibrations described in this paragraph, one might optimistically conclude that $Y_{\rm s}=0.24 \pm 0.01$, while recognizing that the accuracy might well have been overestimated.  Consequent estimates of $\overline{\delta_{\rm{int}} {\rm ln}\gamma_1}$ are illustrated in Figure \ref{fig7}.

\subsection{Other Matters to Consider}
There are  uncertainties in other quantities used in estimations of the chemical composition that still need to be addressed.
Amongst them are the current luminosity [$L_{\odot}$] and the age [$t_\odot$], a matter to
which I have already alluded.  The luminosity is inferred from the
total solar irradiance,  and is currently obtained by assuming that the radiant output is
spherically symmetric \citep[\textit{e.g.}][]{willsonhudsonirradianceluminosity1988Natur}.  That assumption is not strictly correct, as
models of solar-cycle irradiance variation indicate
\citep[\textit{e.g.}][]{L_TSIconfusion1998SoPh, Foukaletalsolarirradiance-luminosity2006Natur, frohlichsolarirradiancereview2011SSRv}; accounting for the
asphericity augments the estimate of $L_\odot$ by some 1.5\,\% or so.  However, there has been
a recent downward revision in  the measured value of the irradiance
from an end-to-end instrumental recalibration \citep{leankoppnewtsi2011GeoRL.3801706K}, which
is partially compensating.  The age [$t_\odot$] is more uncertain.
It is likely to be about the same or only slightly greater than the
age of the oldest meteorites, which seem to lie between $4.563$ and $4.576$ Gy  \citep{amelinmeteoriticages2002Sci, Jacobsenetalmeteoriticages2008E&PSL, Jacobsenetalerratum2009E&PSL, bouvieretalmeteriticages2010NatGe}.   Attempts to calibrate solar models
seismically have yielded values equal to or slightly lower than the
age $t_{\rm S}$ adopted for Christensen-Dalsgaard's model S, namely $4.60$ Gy.  Those calibrations in
which a value of $Z_{\rm s}$ is assumed to be that of Model S yield a
value rather lower than $t_{\rm S}$ \citep{dziembowskietalsolarage1999A&A, 
bonannoschattlpaternosolarage2002A&A, doganboannoJCDsurfaceeffectsandage2010},  
the most recent, based on BiSON data \citep{chaplinetalmodefrequencies2007ApJ...670..872C} being 4.57 Gy;   a
simultaneous calibration of $Z_{\rm s}$ and $t_\odot$ by \citet{GHDOGsolarage2011MNRAS}, also with BiSON data \citep{BiSONfrequenciesbasuetal2007ApJ...655..660B}, has 
yielded precisely $t_{\rm S}$.  Interestingly, the heavy-element abundance
in the latter investigation was calibrated to be $Z_{\rm s}=0.0142$, essentially the same as the 
recent value preferred by \citet{lodderspalmegail_solarZ_2009}, and lying between
the values preferred  by \citet{Asplundetal2009ARA&A}  and 
\citet{Caffauetal2011SolPh}.  However, the model
was calculated with the relative heavy-element abundances of \citet{Grevessenoels1993} 
(neon not being artificially enhanced);  therefore the opacity is too low and the seismic structure cannot
be correct\footnote{Moreover, the initial helium abundance [$Y_0$] of the calibrated model is about 0.250, which is dangerously close to the amount [$Y_{\rm p}$] believed  to have been created by Big-Bang  nucleosynthesis, whose estimated value has been climbing over the last two decades \citep{steigman_BigBangnucleosynthesis_2007ARNPS..57..463S}: the latest estimate is $Y_{\rm p}=0.2478\pm0.0006$ (G. Steigman and M. Pettini, personal communication, 2011).  Subsequent contamination of the interstellar medium by supernovae exacerbates the situation.}.  
Whether that has yielded a superior or an inferior
estimate of $t_\odot$ is unclear \citep{DOG2012Fujihara}.  The calibration relies
partially on estimating $Y_{\rm s}$ seismologically from the oscillatory
component of the low-degree frequency distribution caused by the
acoustic glitch associated with the depression of $\gamma_1$ in the
helium ionization zones \citep{HGseconddiff2007MNRAS}.  To accomplish that
estimate it was assumed that hydrostatic support is solely a balance
between pressure gradient and gravity, as is usual: magnetic stresses  and the centrifugal force were ignored. (It was also assumed that the Sun
is spherically symmetric, as in other calibrations, which we know is not strictly correct \citep[\textit{e.g.}][]{asphericityvariation2002AdSpR..29.1889G,helioseismicsolarcyclesensing2006AdSpR.38.845K,emilioetal2007ApJ...660L.161E,fivianetal2008,kuhnetalsolarshapeScience2012};  by how much the structural asymmetry contaminates the assumed relation between the degree-dependence of the seismic frequencies and the locations of their lower turning points, of crucial importance to the structural calibrations, is yet to be ascertained.)   

The validity of ignoring magnetic stress has been implicitly questioned by \citet{BM2004ApJ.617L.155B} and \citet{verneretalsolarcyclevariation2006ApJ}, who reported a 10\,\%
solar-cycle variation in the glitch signal.  That
variation is huge, and, if correct, cannot possibly result from a
temporal abundance variation. It has been pointed out that were the glitch variation to be magnetic, an intensity variation of
some ten or so  Tesla in the second He ionization zone would be
implied \citep{DOGNaxosconf2006ESASP.617E.1G}.  Moreover,  it would introduce an error in the two-parameter age calibration of similar relative magnitude \citep{GHDOGsolarage2011MNRAS}.   However, \citet{JCDmariorempelMJTovershoot2011MNRAS}  
have recently failed to detect a
change of such magnitude, although they admit that further analysis is
necessary in order to be sure.  Meanwhile, \citet{basuetalsolarcyclevariation2010} maintain that 
the change does occur.  
The whole issue is evidently very important, and we would like to know
the answer.  It may be some time before we do.

\section{What of Importance is There Left to be Learnt?}

Helioseismology was amazingly successful in the early days.  It
demonstrated that the resolution of the solar neutrino problem was not
to be found in adjustments to parameters of a standard solar model,
and later, as precision increased, that the resolution must be sought
in nuclear or particle physics.  As we all now know, the matter is
resolved by neutrino transitions.   Helioseismology has also provided
estimates of the solar helium abundance in the convection zone, perhaps not as accurately as we
would like, yet which, granted gravitational settling,  are at least not seriously at variance with 
Big-Bang nucleosynthesis. Coupled with that is a precise measure of the
location of the base of the convection zone \citep{jcddogmjtconvzonedepth1991ApJ...378..413C, basuantia1997MNRAS.287..189B}, which has been used
extensively as a simple diagnostic for calibrating theoretical solar
models \citep[\textit{e.g.}][]{turckchiezeetalreview1993PhR...230...57T, Bahcallpinsonneaultbasu2001ApJ, 
guziketalsolarcontamination2005ApJ}, and is important for defining the boundary conditions in
numerical simulations of the convection zone.  In addition, we know that
the quadrupole moment [$J_2$] of the exterior gravitational
equiptentials is about $2.2 \times 10^{-7}$ \citep[\textit{e.g.}][]{schouetalrotation1998ApJ, antiachitregoughsolarKE2008A&A},  contributing to the precession of
planetary orbits, particularly that of Mercury, by an amount that is
compatible with General Relativity.  The latitudinal variation of the
angular velocity [$\Omega$] in the photosphere persists, approximately, to the base of the
convection zone, beneath which is a thin interface, the tachocline,
and a rigidly rotating interior (admittedly with some doubt about
the rotation of the energy-generating core).  

We seem to have answered the major
outstanding questions accessible to seismology that interest most of
the scientific community.  So is global helioseismology finished?  Is
helioseismology finished?  To be sure, there is the unfinished business
I talked about in the previous section, but it may require only minor
details to sort that out.  It is now often suggested by commentators that the days of true excitement in the subject, and of important discovery, are over.

I dissent vehemently from that view.  Firstly, the issues that I
discussed in the previous section are related to the microphysics of opacity and the equation of
state, which are important to plasma physics and may end up being very 
important also to the study of stars other than the Sun.  More detailed
seismological investigation may be able to unveil, or at least provide
clues to, some fundamental unresolved issues in plasma physics, such as, in the chemical
picture, how the energy difference between the continuum of ``unbound''
electrons and the strongly bound states is determined.  This is
related  to the whole matter of electron screening of charged species,
a reliable consistent quantum-mechanical study of which is still
wanting, and has implications also for thermonuclear reaction
rates \citep[\textit{e.g.}][]{dappenEoS1998SSRv}.  But more obvious is the whole matter of the internal 
macroscopic dynamics of the Sun: large- and medium-scale
angular-momentum transport; material redistribution by meridional flow
in the radiative zone; the pattern of the large-scale meridional flow in the convection zone;
augmentation, distortion, and decay of the magnetic field and its
back-reaction on the flow; the formation, evolution, and final decay of
sunspots and other forms of activity; and how all these conspire to
drive and control the solar cycle, and, perhaps most prominently,
modulate the solar outputs of particles and electromagnetic radiation
that influence the climate on Earth.  Many of these are the object of
ongoing local helioseismological techniques -- time--distance and
seismic holography, and ring 
analysis.  I have not discussed those techniques here.  Ring analysis
is relatively straightforward, and has given us views of horizontal
flow not far beneath the photosphere.  Telechronoseismology has given us a  view of a sunspot, complete with the flow around it:  deep divergent horizontal flow carrying away the excess heat  rising around the obstructing spot and an associated convergent subphotospheric counterflow above \citep{junweietalsunspotstructureI2001ApJ, junweisashasunspotstructureII2003ApJ, agk2009SSRv..144..175K, zks_subsurf_structures2010ApJ...708..304Z}.   Divergent (Evershed) motion has now been detected even closer to the
photosphere  \citep{evershedoutflow2011SoPh.tmp.163Z} -- a critical feature that was
missing from the original seismological analysis, and which caused some onlookers to
harbour grave doubts.  The qualitative picture,  which has made its way onto the well-known
SOI/MDI coffee mug, is now not only physically plausible, but also in accord with photospheric  observations.   A dynamical picture of the Evershed flow is now emerging \citep[\textit{e.g.}][]{WTBTpenumbralstructure2004ApJ...600.1073W,Ilenasashanourievershed2009ApJ...700L.178K,penumbralseaserpent2010ApJ...716L.181K}.  More recent helioseismic observations \citep{sasahtomdecayingsunspots2011arXiv1102.3961K} have revealed the cessation of the deeper subphotospheric convergent flows as sunspots decay, flows which one might presume had previously held the spots intact.   However, different seismological analyses are not yet all entirely consistent \citep[\textit{e.g.}][]{sunspotseismology2009SSRv..144..249G,HHT2009ApJ...698.1749H}, so the picture is not wholly secure.  Also, more extensive observations will be needed for understanding the entire life-cycle of the spots.


Another matter of importance for understanding the global internal
dynamics is the structure of the tachocline and its associated meridional flow, and how
that influences the convection above.  It is extremely difficult to detect deep meridional flow seismologically, because the frequency perturbations are small; however,  some progress 
appears to be possible with the use of eigenfunction distortions 
\citep{schouetalmeridionalflow2009SPD....40.0705S, DOGBWHmeridionalflow2010ApJ...714..960G}, and 
\citet{antiachitregoughmagneticrotation2012mnras} are trying to advance simplified 
seismologically constrained dynamical arguments.  If the meridional tachocline
flow is downwelling near the Equator and the Poles, and upwelling near
the latitude of zero tachocline shear, as \citet{easjpztach1992A&A} and
\citet{dogmem1998Nature} have argued, how does that flow connect to
the general meridional flow in the convection zone?  Or is that flow so slow
that it is simply swept aside unnoticed by an independent
Reynolds-stress-driven circulation?  Does the upwelling dredge up a
primordial magnetic field from beneath, and if so, does the shearing
of that field in the tachocline react back on the rotation to produce
a mid-latitude shear-free region?  Does the dredged field stoke a
(non)dynamo in the convection zone \citep[\textit{cf.}][]{benjennicdogstoking}?
Does the equatorward tachocline flow in the polar regions and the
poleward flow in the equatorial regions advect the periphery of the
predominantly dipole remnant field in the radiative interior to align
with the upwelling flow-convergence zone \citep{DOG2012GApFD}, or do the torques from the
rotational shear dominate the dynamics, as Toby Wood and Michael McIntyre (in preparation) quite plausibly
presume?  Is such a putative inclined dipole responsible for the
active longitudes? --  and the emergence of sunspots?  Most previous dynamical studies have presumed the tachocline flow to be essentially steady; the evidence presented recently by 
\citet{antiachitregoughmagneticrotation2012mnras} suggests that it varies with the solar cycle.  So these issues will need to be readdressed in a new light.  The rise of
sunspots through the convection is now becoming accessible to seismic
observation \citep{stathisetalemergingsunspots2011Sci}.  There are also questions related to structural
changes associated with the cycle, and how they are related to the
variations in the Sun's outputs.
And what of the rotation of the solar core, and the associated
circulation, if $\Omega$ differs substantially from the angular
velocity of the surrounding envelope?  That is likely to influence our
ideas about other stars.  

It must be appreciated that the Sun is an
important benchmark for the whole of stellar physics, and
helioseismology for asteroseismology.
Finding answers to many of these questions is likely to be assisted by
new helioseismological findings, although seismology alone will not be
enough.  Answers will not come easily -- the easy questions were
answered quickly (although not necessarily easily) in the early days.  But the rewards from answering
the new questions -- designing subtle, hardly detectable, seismic
diagnostics and developing techniques to analyse them -- are
potentially great.  Their pursuit is to a large degree the task of the
new generation of young seismologists.

\section{Acknowledgements}
I am grateful to J\o rgen Christensen-Dalsgaard, Werner D\"appen, G\"unter Houdek, and Sasha Kosovichev for interesting discussion.   I thank Jeannette Gilbert and Paula Younger for typing the first draft of this  paper, G\"unter Houdek for providing Figure \ref{fig4}, and Amanda Smith for her help in producing Figures \ref{fig1} and \ref{fig6}.
I thank the Leverhulme Trust for an Emeritus Fellowship, and P.H. Scherrer for support from HMI NASA contract NAS5-02139.  This article has benefitted from comments on the first draft by a referee who raised the matter of the reliability of the equation of state.


\bibliographystyle{spr-mp-sola}

\bibliography{references}

\end{article}
\end{document}